\documentclass[fleqn,10pt]{wlscirep}
\usepackage{upgreek}
\usepackage{ragged2e}
\usepackage[T1]{fontenc}
\usepackage[english]{babel}
\usepackage{lipsum}
\usepackage{blindtext}
\usepackage{ulem}
\usepackage{threeparttable}
\usepackage{subcaption}

\newcommand{\tidx}[1]{_{\text{{#1}}}}
\newcommand{\OverSqrtHz}{/$\sqrt{\text{Hz}}$}
\title{Characterizing Timing Parameters in Commercial SERF-OPM Multichannel Systems for Biomagnetic Field Sensing}
\author[1,2,*,+]{Eric Elzenheimer}
\author[2,+]{Hartmut Matz}
\author[3,2]{Jan Zerfowski}
\author[2]{Paul Anders}
\author[1]{Michael H\"oft}
\author[1]{Robert Rieger}
\author[3]{Surjo R. Soekadar}\author[4]{Stephen Robinson}
\author[2,+]{Silvia Knappe-Gr\"uneberg}
\affil[1]{Department of Electrical and Information Engineering, Kiel University, Kiel, Germany (e-mail: ee@tf.uni-kiel.de)}
\affil[2]{Department 8.2 Biosignals, Physikalisch-Technische Bundesanstalt (PTB), Abbestraße 2-12, Berlin, Germany}
\affil[3]{Clinical Neurotechnology Lab, Charité -- Universitätsmedizin Berlin, Berlin, Germany}
\affil[4]{MEG Core Facility, NIMH, NIH, Bethesda, MD, USA}
\affil[+]{these authors contributed equally to this work}

\begin{abstract}
Commercial multichannel Spin-Exchange Relaxation-Free Optically Pumped Magnetometers (SERF-OPMs) with intrinsic noise levels in the low femtotesla range are emerging as viable alternatives to Superconducting Quantum Interference Device (SQUID)-based systems, especially for Magnetoencephalography (MEG). Typical SERF-OPM performance limitations and error sources necessitate countermeasures that must be anticipated by the manufacturer and integrated into the system. This enhances the system's expressiveness and comparability for diagnostic applications, thereby supporting acceptance by medical professionals. However, the timing characteristics of SERF-OPM multichannel systems, particularly in terms of time delay, group delay, corresponding intra-channel spread, and settling time, remain poorly characterized. In this initial study, the required timing parameters are defined, quantified, and interesting aspects for system users are discussed. All investigations are conducted using the previously introduced magnetometer test bench (DALAC), operated within the BMSR-2.1 at the Physikalisch-Technische Bundesanstalt (PTB, Berlin, Germany), on four commercial multichannel systems: the fully integrated 3D Neuro-1 and the predecessor 2D QZFM Gen.2 units (QuSpin Inc., Louisville, CO, USA), as well as the latest update of the fully integrated 2.5D HEDscan and the 1D predecessor FLv2 system (FieldLine Inc., Boulder, CO, USA). As a result, the frequency-dependent parameters, presented as plots versus frequency, reveal a time delay of 1-10\,ms with an intra-channel spread of up to $\pm$1\,ms in the 20-140\,Hz frequency range, a group delay variation of 1-15\,ms, and a settling time between 2 and 55\,ms based on a 5\,\% criterion. These results indicate that the two manufacturers apply distinct optimization strategies in their system designs. While QuSpin Inc. appears to manage and minimize intra-channel spread by statistical parameter modeling in the 20-140\,Hz frequency range, FieldLine Inc. focuses on achieving broader bandwidth through per-device calibration. The findings provide initial benchmarks for assessing the timing-critical performance of multichannel SERF-OPM systems and underscore the importance of calibration strategies in preserving waveforem integrity, an essential requirement not only for biomagnetic diagnostics but also for timing-sensitive applications such as stimulation-evoked measurements (e.g., evoked Magnetomyography, Magnetoneurography), Brain-Computer Interfaces (BCI), or closed-loop neuromodulation.
\end{abstract}
\begin{document}
\flushbottom
\maketitle
\thispagestyle{empty}
\section*{Introduction}
The detection of ultra-low biomagnetic fields using multichannel magnetometer systems (MCS) continues to drive rapid research in magnetoencephalography (MEG) \cite{Schofield.2024, Alem.2023, Tatsuoka.2023, Schwindt.2022, Brookes.2022}, brain-computer interfaces (BCI) \cite{Tan.2025,Zerfowski.2021}, fetal magnetocardiography (fMCG) \cite{Strand.2019, Batie.2018, Alem.2015, Sheridan.2010, Vrba.2004, Preissl.2004, Blum.1985}, magnetospinography (MSG) \cite{Spedden.2024, Mardell.2024}, and magnetomyography (MMG) for electrically evoked muscle responses \cite{Nordenstrom.2024, Elzenheimer.2020, Semeia.2022}. Among the available magnetometer technologies, Spin-Exchange Relaxation-Free Optically Pumped Magnetometers \cite{Romalis.2022} (SERF-OPMs) and paramagnetic resonance-based $^4$He-OPMs \cite{ Fourcault.2021, Spedden.2024, Mardell.2024,Bonnet.2025} are currently the only systems that achieve the required femtotesla noise level without cryogenic cooling while operating within the bandwidth of interest. They offer a small footprint, flexible sensor positioning and are capable of measuring the magnetic vector components with Signal-to-Noise ratios (SNRs) > 60\,dB. Together with advances in statistical digital signal processing and computing power, these features pave the way for wearable, mobile, and cost-effective platforms that operate in magnetically shielded rooms (MSRs) and are compatible with subject-specific helmets. In addition, SERF-OPMs are commercially available in sufficient quantities to meet the growing demand of medical research centers worldwide, making them a potential sensing technology for biomagnetic research. Their diagnostic and clinical relevance has not yet been demonstrated, but progress in neuroscience is apparent: more than 100 institutes already operate (SQUID-based) MEG systems \cite{frohlich.2023}. For clinical certification, however, reproducibility, system-independent results \cite{Sun.2024,Bonnet.2025,Tanner.2025}, and cross-platform comparability for MEG still need to be proven. Consequently, challenges such as reproducibility, lifetime, and cross-system consistency must be addressed to ensure accurate and reliable measurements with SERF-OPM multichannel systems.\\\vspace{-3mm}

Important, yet often overlooked and underestimated parameters in multichannel SERF-OPM systems, are the \textbf{time delay ($t\tidx{d}$), group delay ($t\tidx{g}$), and settling time ($t\tidx{set}$)}. Time delay is the relative time difference between a sinusoidal input (monofrequency signal), defined by the to-be-measured magnetic signal, and the response of the sensor, recorded synchronously by the same data acquisition; it can be converted to phase delay ($\Delta\varphi$) and vice versa. The group delay captures the frequency-dependent variation in signal propagation, which is particularly relevant for broadband signals and multi-frequency signals. Settling time indicates the response to a sudden change in the $B$-field, i.e., the time it takes for a sensor to "lock onto" a signal after a disturbance or excitation change. During this time interval, the recorded data contain no field information. For medical applications, time delay, group delay, and settling time are typically expressed in milliseconds (ms). Timing performance in magnetometer systems originate from mainly six key contributing factors: (I) spread of time delay between individual channels, (II) a temporal offset in the collective sensor response relative to the to-be-measured test field, (III) group delay dispersing signal frequency groups, wave envelopes or evoked signal bursts, (IV) settling time, (V) delays introduced by Digital-to-Analog (DA) or Analog-to-Digital (AD) conversion, including lock-in processing and demodulation, and (VI) network delay as an additional latency between the data acquisition (DAQ) system and the recording computer, especially relevant for BCI applications or closed-loop neuromodulation applications when the DAQ is not the master time base. The effects outlined in (IV) and (V) are beyond the scope of this study and are therefore not investigated here.\\\vspace{-3mm}

In this study, timing characteristic is analyzed across four commercial multichannel SERF-OPM MEG systems: a) the Neuro-1 (Diva) fully integrated, closed-loop (CL) triaxial multichannel system; b) its predecessor, QZFM Gen. 2, based on several open-loop (OL) dual-axes sensor units manufactured by QuSpin Inc. (QuSpin Inc., Louisville, CO, USA); c) FieldLine Inc.’s (Boulder, CO, USA) fully-integrated CL HEDscan MCS (latest update: April 2025); and d) the earlier single-axis predecessor CL/OL FLv2 system. The commercially available and technically promising \textsuperscript{4}He-OPM-based 3D-vector magnetometer system from MAG4Health \cite{Fourcault.2021, Bonnet.2025, Spedden.2024, Mardell.2024} was not accessible during the investigation period and is therefore not included in this study. OPM Sensors can be operated usually in single-, dual-, or triaxial mode, referring to the number of active sensitive axes per sensor head. The flux-locked loop (FLL) mode is referred to as Closed Loop (CL), whereas Open Loop (OL) denotes operation without active field cancellation. This study investigates only a subset of the available operating modes.\\\vspace{-3mm}

The manuscript is structured as follows: The first section provides \textit{Motivation and Background} for time delay studies in the context of biomedical applications. All measurements were conducted inside the BMSR-2.1 (Berlin Magnetically Shielded Room\cite{EE_SK_Mag_Sensors.2025}) using homogeneous, well-characterized magnetic test fields of varying amplitudes and frequencies, where the term "amplitude" refers to the peak value of a sinusoidal signal in the mathematical sense. Test bench, data acquisition, and data analysis are described in section \textit{Experimental Setup}, including definition and error estimation of the different key values. A detailed report on the experimental setup and environmental conditions is provided in a previous publication \cite{EE_SK_Mag_Sensors.2025}. The results of the four MCSs are presented in the section \textit{Results} and interpreted in the section \textit{Discussion} with respect to biomedical applications. Finally, section \textit{Conclusion} culminates the paper and addresses the present limitations of timing issues. 

\section*{Background and Motivation}
\label{sec:background}
Multichannel SERF-OPM systems offer three key advantages over SQUID systems that are frequently utilized: non-cryogenic operation, a reduced source-to-sensor distance, and flexible sensor positioning tailored to the specific experimental task. Because SERF-OPMs employ modulation/demodulation techniques, they exhibit more pronounced time- and group-delay characteristics than unmodulated sensors. Particular attention is required to prevent aliasing effects that could compromise signal integrity. The following list establishes the connection between key parameters examined in this study and their significance for specific biomedical applications:

\begin{itemize}
\item (I) Minimum intra-channel spread of time delay:\\
Accurate \textbf{high-resolution dipole-source localization} in MEG requires minimal intra-channel spread in time delay, gain, sensing direction, and synchronous signal processing across all channels in a multichannel sensor arrangement in order to solve the inverse problem. Any temporal misalignment may result in outlier channels that must be corrected or excluded from the model, thereby reducing the effective number of usable channels and degrading overall localization quality and precision. In techniques such as \textbf{trilateration}, even minor discrepancies in phase/timing between channels and sensing directions can result in significant spatial errors \cite{Zekavat.2019}. Similarly, \textbf{gradiometer} configurations require precise time alignment across all channels to suppress external interference effectively \cite{Seymour.2022}, e.g., an electronic multi-channel gradiometer array can achieve a CMRR of about 100 at 30\,Hz when the only contributing uncertainty is time delay variation of IQR\{$t\tidx{d}$\}=100\,$\upmu$s with a sensor bandwidth of about 360\,Hz (all other parameter variations are assumed insignificant). This prediction implies the difficulties associated with electronic gradiometer applications for highly variable OPM sensors. The SSP (Signal Space Projection) algorithm can realize significantly higher noise cancellation efficiencies.

\item (I) and (II) time delay as the collective sensor response relative to the to-be-measured signal:\\
\textbf{Stimulation-evoked responses}, such as latency tracking of the N20 component or the determination of \textbf{Nerve Conduction Velocity} (NCV) by Magnetomyography (MMG), are highly sensitive to even minimal time delays. For instance, a system-induced delay in the microsecond range can introduce significant inaccuracies in the interpretation of stimulation-evoked signals, potentially affecting both diagnostic outcomes and the validity of pilot research studies, particularly when aiming to detect the result of a physiological delay spread (temporal dispersion \cite{WSM.1999}) inherent to individual nerve fibers \cite{Elzenheimer.2020}. Segmental analysis, a standard technique in Electrodiagnosis (EDX) within neurology, addresses this issue by measuring relative delays between distal and proximal stimulation sites with a supported kilohertz-range measurement bandwidth \cite{Elzenheimer.2018, Kimura.2013}. However, this approach is not feasible by default in magnetic measurements, as the close proximity between the distal stimulation site and the highly sensitive magnetometer can lead to sensor saturation or loss of lock. This is primarily due to the stimulation current (mA range) required to satisfy full physiological activation (supramaximal stimulation) of all nerve fibers and to ensure subject-independent comparability of the results \cite{Elzenheimer.2020}. Similarly, the accurate determination of \textbf{Muscle Fiber Conduction Velocity} (MFCV) using multiple SERF-OPM sensors for segmental \textbf{assessment of spontaneous activity} \cite{Baier.2025} critically depends on precise synchronization to minimize intra-channel delay variation. Even when correlation-based methods are employed for delay estimation, such variability can markedly affect the accuracy of temporal signal alignment and, consequently, compromise the reliability of MFCV computations. Finally, the implementation of \textbf{global feedback systems} for active quasi-static field stabilization imposes strict requirements on minimal loop delay and low inter-channel variation \cite{Brookes.2021, Holmes.2023}.

\item (III) Group delay:\\ 
Group delay, the frequency derivative of phase, governs how the information-carrying envelope of a broadband signal propagates through the sensor and associated electronics. Unlike single-frequency phase delay during steady-state conditions, group delay is strictly causal and, therefore, cannot be negative. Ideally, group delay remains constant across both frequency and amplitude; any deviation introduces non-linear distortion that broadens fast transients, alters relative timing, and ultimately distorts the waveform. Such variability compromises \textbf{signal integrity} (or “fidelity”), degrading broadband signals. Fig.\,\ref{fig:sim_mcg_example} will demonstrate the primary influence of group delay on a Magnetocardiography (MCG) signal, highlighting its effect on exaggerated magnetometer values. This phenomenon is certainly not limited to the MCG and can be applied to other applications where the group delays of current OPM MCS systems are the limiting factor. For example, Magnetorelaxometry (MRX) signals, whose frequency content depends on the relaxation dynamics of the selected magnetic nanoparticles (MNP), with faster decays corresponding to higher frequency components, a non-constant group delay can distort the sharp initial decay of the relaxation curve. This temporal smearing can lead to biased or inaccurate estimation of common relaxation parameters \cite{Jaufenthaler.2022}, thereby affecting the analytical interpretation of MRX-based analyses.

\item (IV) Settling time:\\
Adequate settling time is crucial when recording repeated \textbf{single-event signals}. If the sensor has not fully re-established its operating point, signal amplitudes are attenuated, and frequency-dependent features distorted. This limitation is most acute for stimulation-evoked responses, which have micro- to millisecond pulses delivered at high repetition rates for signal averaging. \textbf{Transient interferences, sudden disturbances}, can unlock the sensor's flux feedback loop and require settling towards a new operation point by activation of the self-resonance. Information lost during this time interval cannot be recovered by FFT-based post-processing or adaptive gradient-based algorithms (e.g., Least Mean-Squares Algorithms, LMS). The same constraint applies to interference nominally outside the specified bandwidth, such as power-line harmonics or broadband magnetic fields originating from medical devices, including deep-brain stimulators \cite{Yalaz.2024}.
\end{itemize}

Another critical factor is network delay, which can significantly impact performance in real-time and time-sensitive measurement applications, such as \textbf{Brain-Computer Interfaces}, where timing is essential for accurately interpreting neural responses \cite{Zerfowski.2021}, establishing precise conditioning of neural activity, and maintaining perceptually continuous feedback. Although the timing requirements for BCI depend on the particular application, providing perceptually continuous feedback to the user requires total delays (sensor, acquisition, network, data processing, and feedback delay) below 50\,ms \cite{Edelman.2024}. To perform phase-targeted stimulation (e.g., in the gamma band), the feedback must be computed and applied with even shorter delays.

A single-channel MCG\, \cite{Elzenheimer.2021} recorded using the PTB 304-channel SQUID Vector Magnetometer System (VMS-304\, \cite{Bechstein.2006, Schnabel.2004}), is used to illustrate how magnetometer system characteristics can influence signal integrity, specifically, the signal shape. Fig.\,\ref{fig:sim_mcg_example} shows the simulated impact of introducing exaggerated values of time delay, group delay, and settling time on the recorded SQUID MCG signal:
\begin{figure*}[!ht]
    \centering        
    \includegraphics[width=0.75\textwidth]{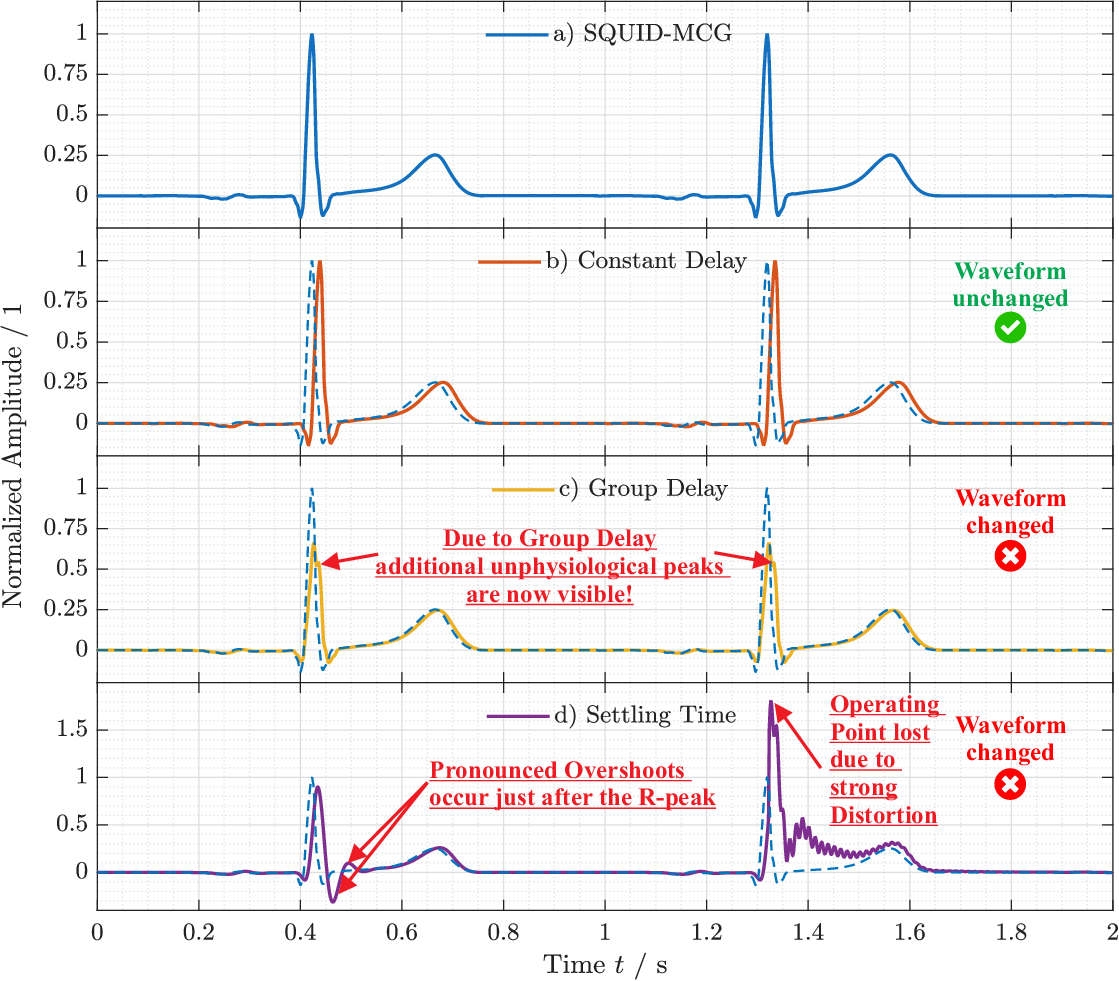}
    \caption{Simulated effects on a single-channel recorded SQUID-MCG signal: a) original waveform, b) introduced constant time delay of 15.6\,ms, c) introduced frequency-dependent group delay rate of 7\,ms per Hz, and d) introduced a settling time of $\tau$ = 98\,ms, combined with an operating point loss triggered by a large field distortion event at the second heartbeat.} 
    \label{fig:sim_mcg_example}
\end{figure*}
a) The original SQUID-recorded MCG waveform, which serves as a reference trace for subsequent simulations, has an R-Peak around 70\,pT. b) Introduction of a constant 15.6\,ms delay (solid copper red line), shown against the original SQUID MCG (dashed line). A uniform shift of the entire waveform is observed without altering the signal’s content. c) Showcasing a frequency-dependent group delay by choosing a frequency-dependent time delay rate of 7\,ms per Hz. Higher-frequency components, such as those around the QRS complex, experience a pronounced delay, causing R-peak broadening and reduced Q- and S-amplitudes. The P-wave is only slightly shifted due to its lower-frequency content. Overall, the waveform’s characteristics are significantly compromised. d) Simulation of a 98\,ms settling time ($\tau = 98$\,ms, cf. subsection \textit{Settling Time}), primarily affects the QRS complex, resulting in a pronounced overshoot of the original signal near the R-peak (cf. S-complex and ST segment). The signal also fails to reach the true Q-, R-, and S-wave amplitudes in time, altering the actual waveform and compromising overall signal integrity. In addition, a simulated distortion event during the second heartbeat causes the OPM operating point to be lost. Recovery takes time, which affects the signal shape, preventing recording and hiding crucial diagnostic information. These simulations underscore the critical need to identify and, if possible, accurately calibrate a magnetometer system to preserve the waveform and avoid potential misinterpretation or false indication of pathological conditions. The distortion parameters are chosen to demonstrate the specific effects on hard beat and all the above real measured values.

\section*{Experimental Setup and Data Analysis}
\label{sec:setup}
In total, four commercial multichannel SERF-OPM systems are investigated, with intrinsic spectral noise densities $\sqrt{S\tidx{B,int}}$ levels well below 50\,fT\OverSqrtHz: a) The Neuro-1 ("Diva", delivered Dec. 2023) from QuSpin Inc. is stated here as state-of-the-art system, including all newly developed features, like 3D-channels per sensor head in flux-locked-loop operation, a bandwidth of about 150\,Hz limited by the DAQ (sensors operating at higher bandwidth), with calibrated time delay between the channels, cross axis projection error (CAPE) compensation and integrated as multichannel system with DAQ and additional auxiliary channels for evoked signal synchronization. b) The latest update (April\,2025) of FieldLine's HEDscan system with dual-axis measurement capability, 2.5D-FLL operation for CAPE compensation in triaxial mode, and individually phase-, bandwidth-, and amplitude-calibrated sensor channels per head. DAQ, data processing, and software are optimized to record MEG and support BCI applications, which require short latencies between detected brain events and corresponding actuator reactions. c) The predecessor MCS FieldLine v2-system with single-axis CL or OL measurement capability per head and individual phase and amplitude transfer function calibration. Neuro-1 and both FieldLine systems utilize a common DAQ system with sample-synchronized auxiliary channels but slightly different specifications, and both offer a proprietary real-time API for management and data access \cite{quspin2024neuro1, FieldLine.2025}. d) Finally, the well-known and established dual-axis QZFM Gen.2 sensors from QuSpin are included, operating in OL mode, demanding an external DAQ system for signal recording, and representing the origin of multichannel OPM systems. Key evaluation parameters for all four systems are summarized in Tab.\,\ref{tab:overview_systems}. It should be noted that the software version must be explicitly reported for each system. As vendors use their own, often non-transparent, notation schemes, this information is essential to ensure traceability, reproducibility, and cross-system comparability in research.\\\vspace{-3mm}

\begin{table}[b!]
\centering
\caption{Overview of multichannel SERF-OPM systems under test, capable of measuring one, two, or three components of the magnetic field. Each column details the manufacturer, number of magnetometer heads passing the evaluation, number of sensitive axes, software version, feedback condition, head size, standoff distance, CAPE compensation capability, and DAQ system with sampling frequency $f\tidx{s}$, including file format.}
\begin{tabular}{|l|c|c|c|c|}
   \hline
   & & & & \\[-2ex]
	Type & Neuro-1\cite{Schofield.2024} & HEDscan \cite{FieldLine.2025} & FLv2\cite{Alem.2023} & QZFM Gen.2\cite{QuSpin.2022} \\[0.25\normalbaselineskip]
   \hline
   & & & & \\[-2ex]
   Manufacturer & QuSpin Inc. & FieldLine Inc. & FieldLine Inc. & QuSpin Inc. \\[0.15\normalbaselineskip]
   \# of Heads passing the test & 19
   / 21 & 18 / 24 & 17 / 17 & 13 / 16 \\[0.15\normalbaselineskip]
    \,\,\,\,\,Sensitive Axes & 3 & 2.5\,\textsuperscript{\textbf{i)}} & 1 & 2 \\[0.15\normalbaselineskip]
    Software Version & N1\_DAQ\_V3\_53 & hedscan-0.10.12-g64fb6e7 & v1.6.19 & QZFM UI V9.0.23 \\[0.15\normalbaselineskip]
    Feedback Condition & CL & CL & CL, OL & OL \\[0.15\normalbaselineskip]
    Head Size / mm$^3$ & 12.4 x 16.6 x 24.4 & 13 x 15 x 35 & 13 x 15 x 30 & 12.4 x 16.6 x 24.4 \\[0.15\normalbaselineskip]
    \textit{Standoff\,\textsuperscript{ \textbf{ii)}} / mm }& \textit{6.2} & \textit{5} & \textit{5} &  \textit{6.5} \\[0.15\normalbaselineskip]
    CAPE compensation & yes & in 2.5D-mode & no & no \\[0.15\normalbaselineskip]
    DAQ & QuSpin Inc. & FieldLine Inc. & FieldLine Inc. & non\,\textsuperscript{\textbf{iii)}} \\[0.15\normalbaselineskip]
    \,\,\,\, Sampling Frequency / Hz & 375, 750, 1500\,\textsuperscript{ \textbf{iV)}} & 1000 & 1000 &  31..32000 \\[0.15\normalbaselineskip]
    \,\,\,\, File Format\,\textsuperscript{\textbf{v)}} & tdms, lvm & fif & fif & 32-bit float \\[0.15\normalbaselineskip]
    \hline
    \end{tabular}
    \begin{tablenotes}
     \item[i)] \textsuperscript{\textbf{i)}} During assessment, FieldLine Inc. provided a beta version for dual and triaxial CL mode by software update in April 2025. The third (X-) axis is used for CAPE compensation only, with a bandwidth below 10 Hz.
     \item[ii)] \textsuperscript{\textbf{ii)}} Standoff: Distance from the center of the vapor cell to the exterior of the housing along the main axis, given by manufacturers.
     \item[iii)] \textsuperscript{\textbf{iii)}} Used PTB owned data acquisition system (LanDAQ).
     \item[iV)] \textsuperscript{\textbf{iv)}} Sampling frequency of 1~500\,Hz is used for all measurements.
     \item[V)] \textsuperscript{\textbf{v)}} Fractal Image Format *.fif, Technical Data Management Streaming *.tdms, LabVIEW Measurement *.lvm.
   \end{tablenotes}
 \label{tab:overview_systems}
\end{table}
All MCS analyzed in this study were initialized according to the recommended procedures before measurement. This includes sensor zeroing and internal calibration procedure prior to each measurement sequence, as well as after opening and closing the MSR door. The internal sensors' operating parameters for HEDscan MCS were remotely tuned to match the ambient conditions of the specific MSR, here BMSR-2.1, which inherently provides low, stable background noise and minimal residual static fields. The other measured MCSs are used as delivered from the manufacturer. Each system comprises between 16 and 24 sensor heads, each equipped with one to three orthogonal sensitive axes for vector magnetic field measurements. The ratio of functional to inspected heads is listed in Tab.\,\ref{tab:overview_systems} and malfunctioning heads are excluded from further analysis. The table also indicates each system’s measurement mode (CL or OL).\\\vspace{-3mm}

\begin{figure*}[ht!]
    \centering           
\includegraphics[width=1\textwidth]{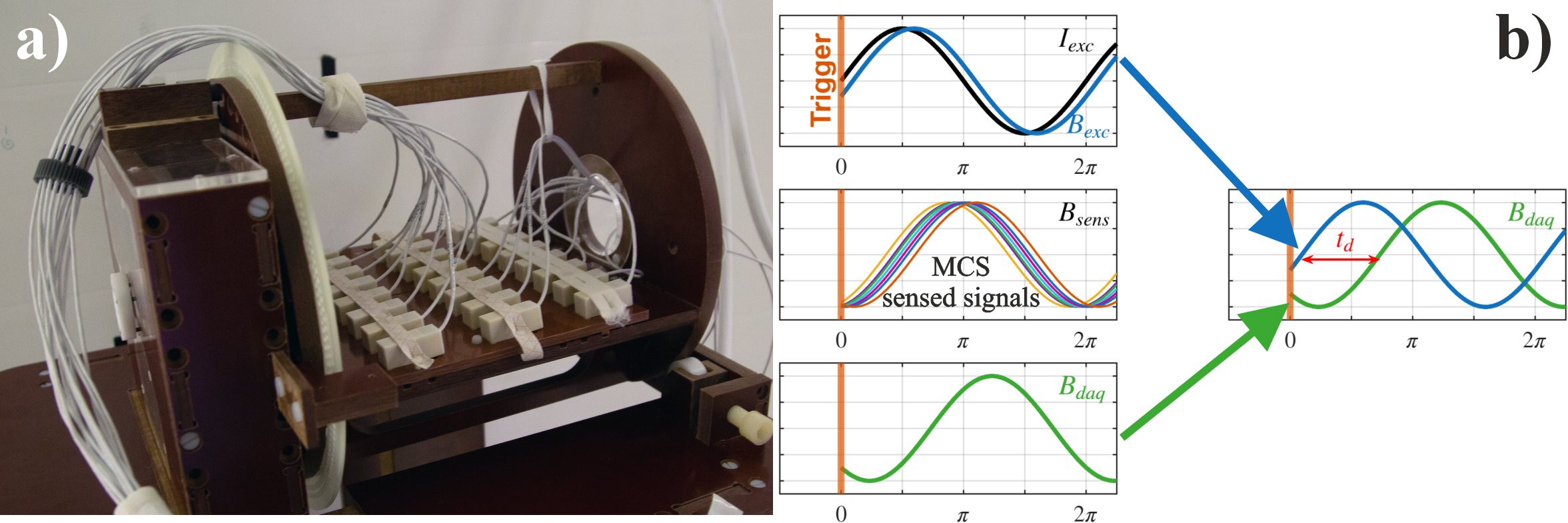}
    \caption{\justifying a) 21 magnetometer heads of HEDscan OPM-MEG System (FieldLine Inc.) are mounted on the DALAC top using a CRP board with precision CNC-milled cut-outs to securely hold each housing. The unit cables are well-fixed to prevent any static B-field change caused by cable movement. b) Scheme of Time-delay evolution exemplified by Neuro-1 MCS: An excitation current applied to the coils generates the magnetic field $B\tidx{exc}$. Eddy currents in the \textmu-metal shielding cause a global phase delay, while each sensor introduces an additional, sensor-specific delay. These are corrected in the data acquisition system during post-processing. The plot depicts the signals, with the Y-axis representing amplitude and the X-axis showing the sine argument in radians. The final time delay $t\tidx{d}$ is subsequently determined.}
    \label{fig:multiple_sensors_at_ dalac_top_and_outside_msr}
\end{figure*} 

The experimental setup, measurement procedure, data acquisition system, and analysis pipeline used in this study have been described in detail in a previous contribution \cite{EE_SK_Mag_Sensors.2025}. The key features are briefly summarized below, and readers are referred to the original publication for more information on the magnetometer test bench, called DALAC (Device ALignment ACtuator). By default, the DALAC ensures precise sensor positioning in all measurements. For this purpose, the sensor heads are aligned on top of this non-magnetic goniometer test bench along three axes with an angular precision of $<$\,0.2\textdegree{} and a spatial precision of $\pm$~2\,mm relative to the MSR center. To enable parallel sensing of the magnetic test field, all sensor heads of a given system were mounted on a cotton-reinforced plastic (CRP) board positioned on the DALAC, as shown in Fig.\,\ref{fig:multiple_sensors_at_ dalac_top_and_outside_msr}. The internal alignment of the laser beam axis within each sensor head was not verified, as this is not practically feasible and can only be achieved by X-ray imaging. A minimum spacing of 30\,mm was maintained between the centers of each unit to minimize crosstalk and prevent thermal coupling between adjacent sensor heads. All measurements were performed inside a large magnetically shielded room \textbf{BMSR-2.1}\cite{EE_SK_Mag_Sensors.2025} (Berlin Magnetically Shielded Room), which provides a shielding factor greater than $>$\,$7 \cdot 10^7$ at 8\,Hz, and a static residual field less than 1.5\,nT. Field drift is limited to below 10\,pT/hour, an important feature to exclude potential phase shift due to CAPE effect \cite{Borna.2022}. The magnetic noise density at 30\,Hz is approximately 0.4\,fT/$\sqrt{\mathrm{Hz}}$ at room center\cite{Storm.2017}.\\\vspace{-3mm} 

A \textbf{homogeneous magnetic test field} is generated by applying an Alternating Current (AC) from a calibrated Keithley 6221 (Tektronix, Beaverton, OR, USA) current source to the integrated large Helmholtz-like coils, with the current precisely recalculated by recording the voltage across a 50\,$\Omega$ precision resistor. The excitation coils are mounted directly and permanently to the inner metal walls of the MSR, resulting in a strong magnetic coupling that provides a stable transfer function. The transfer function has been measured with the 304-VMS SQUID system over a center volume of 30~x~30~x~30\,cm$^3$ and used to determine the time delay between the applied AC current and the generated magnetic AC test field (cf. Fig.\,\ref{fig:multiple_sensors_at_ dalac_top_and_outside_msr}), which is considered and corrected. In this study, the coil system is used to generate low-field amplitudes within a peak amplitude range from $100$\,pT to $10$\,nT, covering a meaningful frequency range from approximately 0.49\,Hz up to 423\,Hz.
\subsection*{Data Acquisition}
All measurements involving reference sensors and QuSpin Gen.~2 SERF-OPMs were recorded using the standard in-house data acquisition system (LanDAQ). In contrast, the Neuro-1 system, FieldLine FLv2, and HEDscan were operated with their respective manufacturer-supplied DAQ systems. A key requirement for accurate time delay analysis was the synchronized recording of two auxiliary input channels, one for the Current Source (CS) and one for the trigger signal, in conjunction with the OPM channels.
\subsubsection*{LanDAQ}
The LanDAQ system is designed and optimized for maximum flexibility, high data integrity, and robust Electromagnetic Compatibility (EMC). It features 304 input channels dedicated to the SQUID-based VMS-304 System, along with 48 differential channels (configurable as single-ended), directly accessible within the MSR, and additional channels for the synchronous recording of environmental conditions. In total, 386 independently sampled channels can be processed and stored in real-time, with sampling rates of up to 32\,kHz. Each channel is equipped with differential input amplifiers, a Programmable Gain Amplifier (PGA), and full-range offset compensation, supporting gain values from 1 to $1 \cdot 10^{4}$ within a gain-bandwidth product of 1.3\,MHz. This enables flexible adaptation to a wide variety of input signal requirements. The system provides 24-bit $\Sigma\Delta$-ADC resolution using a CS5361 chip, with an input voltage range of $\pm~10$\,V and an ADC passband flatness of less than 0.5\,\% below $0.45 \cdot f\tidx{s}$, with output filters having a total group delay of $12 / f\tidx{s}$. Further specifications of the LanDAQ system are detailed in \cite{EE_SK_Mag_Sensors.2025}.\\\vspace{-3mm}

\subsubsection*{DAQs of Commercial SERF-OPM Multichannel Systems}
Although end users frequently request detailed performance specifications for manufacturer-supplied DAQ systems, similar to those available for the LanDAQ, such information is typically not disclosed by most manufacturers. A comparative overview of the DAQ systems is presented in Tab.\,\ref{tab:overview_systems}. The latest commercially available multichannel SERF-OPM MEG systems feature integrated data processing capabilities and user-friendly software interfaces, enabling users to manage multiple magnetometer heads and monitor their operational status. QuSpin Inc. provides a LabVIEW-based software interface for data acquisition with its Neuro-1 system. The system features an integrated architecture that combines QZFM Gen.3 sensors, control electronics, data acquisition hardware, and power management on a single platform, capable of supporting more than 128\,sensors (384\,channels). Each sensor head has an electronic control unit (ECU), positioned inside the MSR, that generates digital output signals. These signals are transmitted to a central motherboard, where they are multiplexed and forwarded via Ethernet to an external data acquisition system (sbRIO9637, National Instruments) located outside the MSR. This DAQ system synchronizes OPM signals with external triggers and transfers data to a host PC through the Ethernet connection \cite{Schofield.2024,quspin2024neuro1}. In addition, QZFM Gen.1 to Gen.3 sensors offer the option of connecting to custom data acquisition systems via analog output ports for signal access. Similarly, FieldLine Inc. offers proprietary software solutions for data visualization, recording, and sensor control, along with real-time access through an Application Programming Interface (API) for its FLv2 and HEDscan system \cite{Alem.2023}, the latter of which can currently be expanded to up to 144 sensor heads \cite{FieldLine.2025}.

\subsection*{Time Delay Estimation}
Several methods, including sine-cosine fitting, cross-correlation, and Fourier transformation, are suitable for estimating the phase or the time delay between an applied signal and the sensor response, with all exhibiting similar results in preliminary evaluations. Each method exhibits specific limitations: SinCosFit is affected by boundary imperfections, cross-correlation is constrained by the sampling frequency, and DFT-based approaches are susceptible to spectral leakage unless ideal conditions are met (such as the usage of a perfectly periodic analysis segment), as demonstrated by Oliveira et al. through Monte Carlo simulations \cite{Oliveira.2024}. Otherwise, the commonly known leakage effect introduces a bias to the phase value. Finally, the SinCosFit method is chosen, being a bit more convenient for large data sets and the maximum 1\,kHz sampling frequency of FieldLine DuTs. Consecutively, sequences with stationary sinusoidal magnetic test fields are applied with different discrete amplitudes $B\tidx{exc}$ and frequencies values, while the test field is applied along the Line-of-Sight (LoS) of the sensitive axis under test. To simplify the analysis, a trigger signal is recorded to mark the start of each signal sequence once the stationary conditions are reached. The applied test field is represented by the current generated by the current source (CS) through the coils, measured across a precision 50\,$\Omega$ resistor and recorded via the auxiliary channels of the DuT’s data acquisition system. The SciPy \texttt{scipy.optimize.leastsq()}\cite{SciPy.2020} function is used to perform a fit to a sine-cosine model function: 
    \begin{equation}
            Y\tidx{out}(t) = C\tidx{0} + C\tidx{1}t + A\tidx{s} \sin\left(2\pi f t\right) + A\tidx{c}\cos\left(2\pi f t\right)
        \label{eq:sincosfit}
    \end{equation}
The offset term $C\tidx{0}$ and the linear term $C\tidx{1}t$ cover a potential, small leftover of the background field compensation and background field drift. In case of a proper zeroing procedure and FLL mode, both terms should be negligible. Details are given elsewhere \cite{EE_SK_Mag_Sensors.2025}. The phase is computed using the two-argument four-quadrant inverse tangent function numpy's \texttt{arctan2(A\textsubscript{s},A\textsubscript{c})}. The time delay $t\tidx{d}$ is calculated from the phase difference ($\Delta \varphi$) between the DuT output signal and the applied current (CS):
    \begin{equation}
t\tidx{d} = -\frac{\Delta \varphi} {f \cdot 360^{\circ}} = \frac{\arctan (\frac{A\tidx{s}}{A\tidx{c}})\tidx{dut} -\arctan (\frac{A\tidx{s}}{A\tidx{c}})\tidx{cs}} {f \cdot 360^{\circ}}
        \label{eq:time_delay}
    \end{equation}
Beware of phase wrapping: once the phase difference nears or crosses $\pm 180^{\circ}$, discontinuities arise. Apply phase-unwrapping algorithms to maintain a continuous phase trace.
\subsubsection*{Statistical Uncertainty in Time Delay Estimation}
\label{subsec:time_delay_error}
Applying Gaussian error propagation to the sine-cosine fit (Eq.\ref{eq:sincosfit}) yields the phase standard uncertainty $u_{\varphi}$ given by:
    \begin{equation}
            u_{\varphi} = \frac{\sqrt{A\tidx{c}^2 \cdot u\tidx{As}^2 + A\tidx{s}^2 \cdot u\tidx{Ac}^2}}{A\tidx{s}^2+A\tidx{c}^2} \,\,\,\, \mathrm{and} \,\,\,\,\,\,\,  u_{\Delta \varphi} = \sqrt{u_{\varphi\mathrm{,DuT}}^2 + u_{\varphi\mathrm{,CS}}^2},
        \label{eq:phase_error}
    \end{equation}
where $A\tidx{s}$ and $A\tidx{c}$ are the sine and cosine coefficients and $u_{A\tidx{s}}$,$u_{A\tidx{c}}$ their uncertainties. The relative uncertainties of the frequencies $f\tidx{DuT}$, $f\tidx{CS}$ are at least two orders of magnitude smaller than the relative amplitude uncertainties of the amplitudes $A\tidx{s}, A\tidx{c}$ and are therefore neglected.
In case of AWGN (Additive White Gaussian Noise), the uncertainty of $t\tidx{d}$ scales with the inverse square root of the number of recorded samples $N$, i.e., measurement time $T\tidx{rec}$ and sampling frequency $f\tidx{s}$ \cite{Kay.2013}. Thus, measurement time is adapted to reach a statistical uncertainty below 50 \textmu s for each sequence. 50 \textmu s is equivalent to 0.009 degrees at the lowest test frequency of 0.49\,Hz. 
\subsubsection*{Systematic Errors}
In general, all measured vector components of a current-induced AC magnetic field are in phase with the applied current, regardless of the sensor's position or the orientation of its sensitive axis in space. In closed, high-permeability magnetic shielded environments such as BMSR-2.1, systematic phase shifts and corresponding time delays can primarily arise from \textbf{eddy current effects} within the inner metallic shield. These effects are particularly relevant at low excitation frequencies within the linear, non-saturated magnetic field range, where the effective skin depth of the shielding material is relatively large. As a result, an additional time delay may be introduced between the current driving the field-generating coils and the actual magnetic test field $B\tidx{exc}$ within the shielded volume. This delay was carefully characterized using the movable 304-VMS system and subsequently used to correct the sensor evaluation data. At a test signal frequency of 1\,Hz, the measured delay between the coil current and the resulting magnetic field at the center of BMSR-2.1 was below  0.28\,ms, decreasing to 69\,\textmu s at the system's cutoff frequency of $f_{\tidx{-3dB}} = 1085$\,Hz. These field characteristics have been validated over several years under varying experimental conditions using the 304-VMS system, consistently showing only minor deviations\footnote{A more detailed discussion of systematic effects related to coil-generated fields in MSRs, specifically in the context of BMSR-2.1, is currently under preparation.}.\\\vspace{-3mm}

Borna et al. \cite{Borna.2022} described a phenomenon known as the Cross-Axis Projection Error (\textbf{CAPE}), in which uncompensated static magnetic fields orthogonal to the sensitive axis induce additional phase shifts and gain- and orientation errors in the SERF-OPM output signals. This sensor technology is particularly susceptible to CAPE due to the intrinsic coupling between angular dependence and phase response. CAPE comes into play in case of near DC fluctuations of the background field or even in a stable, moderate and homogeneous background field, when the sensor heads are tilted by patient movements. The static background field in BMSR-2.1 is exceptionally stable, allowing CAPE effects to be excluded during evaluation measurements. Furthermore, operation in 3D-FLL mode maintains a constant working point and effectively suppresses CAPE. However, residual CAPE-induced phase shifts may still arise from imperfect or non-reproducible field compensation during the OPM initialization (zeroing procedure) or from field changes caused by crosstalk between neighboring sensors, potentially increasing the time delay spread between devices under test.
\subsection*{Group Delay}
Group delay quantifies the time delay experienced by the envelope of a broadband or multi-tone signal as it propagates through a signal processing system. It is inherently positive due to the principle of causality and complies with relativistic constraints, always remaining below the speed of light in vacuum. Group delay is governed by the spectral properties of the input signal and the system’s complex transfer function. In practical systems, especially magnetometers, frequency-dependent group delay can cause signal degradation by introducing unequal delays across spectral components, a phenomenon known in physics as dispersion. This is especially critical in systems where careful reproduction of the original time-domain signal $B(t)$ is essential, such as in biomagnetic recordings (e.g., MCG signals, cf. Fig.\,\ref{fig:sim_mcg_example}~c)). Group delay $t\tidx{g}(f)$ is defined as the negative derivative of the phase response $\Delta \varphi$ with respect to angular frequency $\omega$, and is expressed in seconds (s). Assuming a linear phase model where $-\Delta \varphi (\omega)= t\tidx{d}(f) \cdot \omega$ and $\omega = 2~\pi~f$, the group delay can be expressed as:
\begin{equation}
    t\tidx{g}(f) = \,\,\,\frac{\mathrm{d}(f \cdot \,t\tidx{d}(f))}{\mathrm{d}\,f} = t\tidx{d} + f \cdot \frac{\mathrm{d}\,t\tidx{d}(f)}{\mathrm{d}\,f}
    \label{eq:group_delay}
\end{equation}
Although this definition is conceptually straightforward, estimating the group delay for general stochastic signals remains a challenge. Traditional methods often employ chirp signals, characterized by constant amplitude and continuously varying frequency, to probe the system’s phase response. In this work, however, a more controlled approach is employed: group delay is estimated using a sequence of densely spaced sinusoidal signals with constant amplitude and varying frequencies applied across the sensor’s passband. This method allows for the characterization of all delay contributions introduced by the sensor system, including dead time in FLL mode and delays from digital filtering stages, such as demodulation filters. Furthermore, the time delay associated with each applied frequency is evaluated over multiple signal periods, with precise correction for the inherent delay introduced by the interaction between the test field coil and the \textmu-metal shielding walls of the BMSR-2.1.

In the following, the Neuro-1 system, characterized by a non-constant group delay across its passband, serves as the basis for a functional piecewise group delay analysis. A key assumption is that the group delay profile is smooth and continuous, valid for systems with well-behaved low-order analog or digital filtering characteristics. Two dominant principal frequency regions are identified and modeled to characterize the system's overall time delay behavior $t\tidx{d}\left(f\right)$:
\begin{enumerate}
    \item Low-Frequency Region (LF, $f\approx$ 0.5~-~60\,Hz): Time delay is approximated using an arctangent function to capture the low-frequency region behavior by:
    \begin{equation}
        t\tidx{d,LF}\left(f\right) = \,\,\,b\tidx{1}\cdot\,\, \arctan{\left(j\frac{\,f}{ f\tidx{c,td,LF}}\right)} \,\, + \,\, b\tidx{0}.
        \label{eq:time_delay fit_LF}
    \end{equation}
    Here, $b\tidx{1}$ and $b\tidx{0}$ are fitting coefficients, and $f\tidx{c,td,LF}$ denotes the characteristic edge frequency associated with the time-delay response to capture the low-frequency asymptotics.
    \item Intermediate Region (IM, $f\approx$ 9~-~150\,Hz): In this frequency range, the time delay exhibits a smooth and consistent profile across sensors, indicating the presence of a common underlying transfer function. To capture this behavior, a polynomial of up to 5\textsuperscript{th} order is fitted to the measured time delay data:
    \begin{equation}
        t\tidx{d,IM}(f) = \sum_{n=0}^{n\,\, \leq \,\,5} a\tidx{n}\cdot\left(2~\pi~f\right)^n, 
        \label{eq:time_delay fit_IM}
    \end{equation}
    where $a\tidx{n}$ are the regression-derived polynomial coefficients. The constant term $a\tidx{0}$ introduces an offset, enabling a smoother and more flexible extrapolation throughout the sensor passband.
\end{enumerate}

Adequate frequency overlap between the modeled regions ensures a quasi-continuous approximation of the time delay across the sensor’s entire passband. The resulting piecewise time delay functions are differentiated and combined to yield a continuous group delay characteristic (cf. Eq.\,\ref{eq:group_delay}), while preserving a discontinuity at the cutoff frequency to accurately reflect the physical behavior of the underlying system. This approach effectively captures the sensor’s delay characteristics across its operating range, including limitations introduced by bandwidth constraints and calibration uncertainties. As shown later, the Neuro-1 system notably exhibits a tiled (non-monotonic) delay profile across frequency, underscoring the necessity of combining both arctangent and high-order polynomial models for accurate group delay reconstruction. This procedure is applied to OPM sensors with non-trivial transfer function characteristics. If the polynomial fit demonstrates sufficient accuracy within the primary frequency range of interest (10~–~150\,Hz), the analysis is limited to this interval. The goodness of fit is evaluated to quantify the inherent uncertainty of the estimation process.

\subsection*{Settling Time}
\begin{figure*}[htp!]
    \centering \includegraphics[width=0.7\textwidth]
    {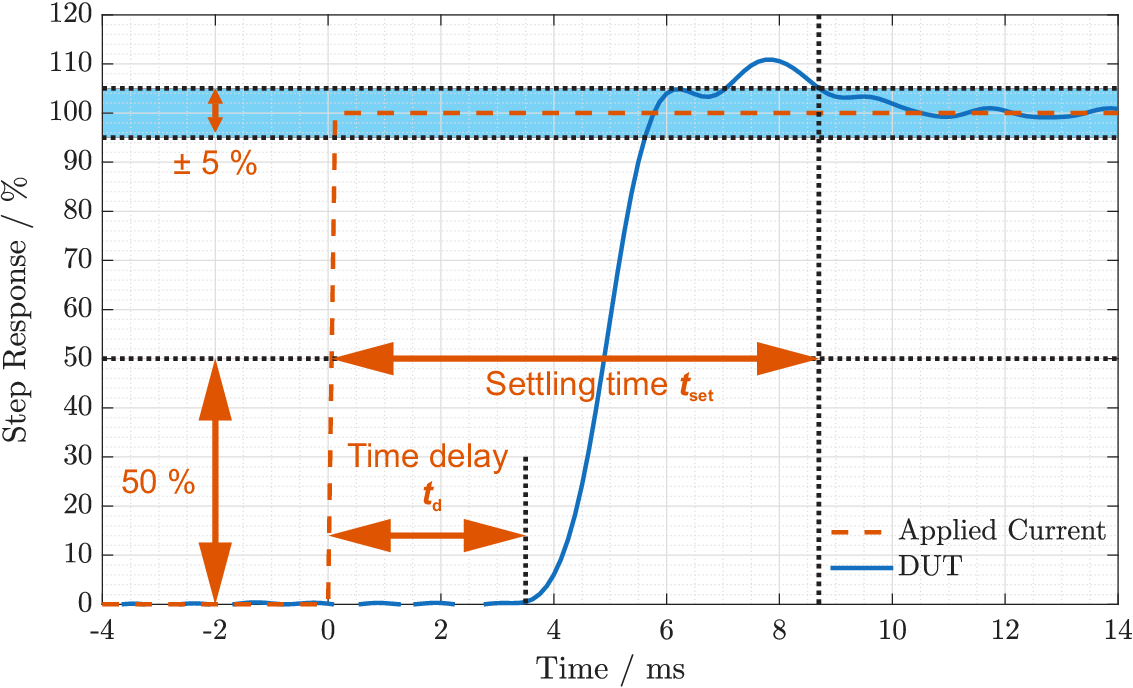}
    \caption{\justifying Illustration of settling $t\tidx{set}$ in response to a magnetic field step change (DC-Change). The dashed red line represents the applied current (normalized to 100\,\%), and the solid blue line shows the response of the device under test, recorded with $f\tidx{s}$~=~8\,kHz. The step onset is defined at $t=0$\,ms, corresponding to the moment when the applied current reaches 50\,\% of its final steady-state value. Settling time $t\tidx{set}$ is measured from this point to the time at which the DuT output remains within a predefined error band, e.g., $\pm$5\,\% around the steady-state value, as indicated by the horizontal back dotted lines. The settling behavior reflects the combined effects of intrinsic time delay, slew rate limitations, and the sensor’s dynamic response characteristics.}
    \label{fig:definition}
\end{figure*}

Settling time is another key benchmark metric that characterizes the dynamic response of a magnetometer system. It represents the sensor’s response to a sudden change in the magnetic field and it quantifies the time required for the output to stabilize or "lock onto" the signal, following a disturbance or a big change in excitation. For evaluation, the step response is triggered by a stepwise change in DC current applied to the test field coils, simulating an abrupt transition in the magnetic field. Formally, settling time is defined as the time interval between the application of a step input (onset) and the point at which the system output remains within a specified fraction of its final steady-state value. The specific threshold is typically adapted to the type of device under test; for example, a 0.1\,\% (1\,\,\textperthousand) criterion is commonly used for data acquisition systems. In this work, error bands of $\pm$5\,\% and $\pm$1\,\% are applied for the evaluation of SERF-OPMs. Fig.\,\ref{fig:definition} illustrates the definition of settling time and highlights its dependence on the interplay of several contributing factors: system time delay, slew rate limitations, and the filter response of the DuT. Signal amplitudes are normalized to the final steady-state value and expressed in \%. The dashed red line represents the applied current, with the step defined at time $t\tidx{0}$, when the recorded current reaches 50\,\% of the full step height. The applied magnetic test field is aligned with the sensitive axis (LoS) of the sensor under test. For QZFM Gen.2 devices operating in open-loop mode, step amplitudes between 100\,pT and 400\,pT are used. Other DuTs, operated in Flux Locked Loop (FLL = CL) mode, are tested with step amplitudes ranging from 100\,pT to 1000\,pT.

\subsubsection*{Uncertainty of Settling Time Estimation}
The uncertainty of settling time measurements depends critically on the timing precision of the applied magnetic field step. To verify that the magnetic test field excitation is sufficiently fast for SERF-OPM evaluation, a SQUID magnetometer is used as a reference sensor. As shown in Fig.\,\ref{fig:settling}~a), the SQUID, sampled at 32\,kHz, demonstrates that the excitation field reaches 50\,\% of its final value within 0.06\,ms and settles within the $\pm$5\,\% error band after 0.9\,ms. This confirms that the step onset is sharp enough to serve as a reliable reference point, and  $t\tidx{0}$ (step onset) can be accurately defined as the time at which the recorded current reaches 50\,\% of the step height. The sampling rate primarily limits the precision of settling time estimation. The lowest sampling rate among the tested systems is 1\,kHz (e.g., HEDscan and FLv2 from FieldLine Inc.), corresponding to a time resolution of $\pm$1\,ms. Therefore, a conservative uncertainty of $\pm$1\,ms is assumed. This includes vertical (amplitude) uncertainty due to quantization error in the measured signal, which affects the detection of when the response enters the error band, as well as horizontal uncertainty. Horizontal (temporal) uncertainty arises from the discrete sampling points when identifying the onset of the current step, $t\tidx{0}$, and the point at which the signal remains within the $\pm$5\,\% error band $t\tidx{set}$. This level of uncertainty is acceptable for the target application of SERF-OPM evaluation, as the typical settling times range from a few milliseconds to several tens of milliseconds, well above the response time of the reference SQUID.

\section*{Results}
\label{sec:results}
The timing parameters of four commercial multichannel SERF-OPM systems were evaluated: Neuro-1 (QuSpin Inc.), HEDscan and FLv2 (FieldLine Inc.), and QZFM Gen.2 (QuSpin Inc.). Table\,\ref{tab:results} summarizes the key results for the investigated systems, including values for time delay (median value for $t\tidx{d}$), group delay (typ. value for $t\tidx{g}$) at $f = 28$\,Hz, and settling time (median value for $t\tidx{set}$) in response to step excitations with amplitudes of 1\,nT (CL) or 400\,pT (OL). The interquartile range (IQR) is used to quantify the intra-channel variation in both time delay and settling time measurements. All measurements were conducted within BMSR-2.1 with the excitation field aligned to each sensor’s body in LoS and averaged over multiple signal periods to reduce statistical uncertainty. A sinusoidal excitation field $B\tidx{exc}$ with constant amplitude was applied at logarithmically spaced frequencies ranging from 0.49\,Hz to 423\,Hz, avoiding frequencies associated with known environmental disturbances, e.g., power-line harmonics and mechanical vibrations within the MSR. Time delay versus frequency data were modeled using appropriate functions, such as arctangent or polynomial fits, to facilitate subsequent group delay analysis (cf. Subsec.~\textit{Group Delay}). A smooth and continuous frequency-dependent behavior of $t\tidx{d}(f)$ and $t\tidx{g}(f)$ was assumed within each system’s -3\,dB-bandwidth. Settling time was defined as previously described (cf. Subsec.~\textit{Settling Time}) and illustrated in Fig.\,\ref{fig:definition}. Settling time measurements contain systematic and statistical uncertainties, especially for very short response times, due to the limited sampling rate of some systems’ data acquisition hardware and minor propagation delays within the MSR environment (cf. Subsec.\,\textit{Uncertainty of Settling Time Estimation}). \\\vspace{-3mm}

\begin{table}[t!]
\centering
\caption{Timing parameters across investigated multichannel SERF-OPM systems, with sensor axes aligned to the applied magnetic test field. Reported values include robust statistical measures (median and interquartile range, IQR) for time delay, group delay, characterizing intra-channel variability.}
\begin{tabular}{|l|c|c|c|c|}
   \hline
   & & & & \\[-2ex]
	Key value & Neuro-1 (19/21) & HEDscan (18/24) & FLv2 (17/17)& QZFM Gen.2 (13/16)\\[0.15\normalbaselineskip]
    & 3D, CL & 2.5D\,\textsuperscript{\textbf{i)}}, CL &  1D, CL \& OL& 2D, OL\\[0.25\normalbaselineskip]
   \hline
   & & & & \\[-2ex]
   Time delay ($f$~=~28\,Hz) / ms &  &  &  & \\[0.15\normalbaselineskip]
   \,\,\, Median & X: 9.3, Y: 8.7, Z: 8.7 & Y: 1.9, Z: 1.8 & Z\textsubscript{CL}: 1.2, Z\textsubscript{OL}: 2.3 & Y: 4.0, Z: 4.0 \\[0.15\normalbaselineskip]
   \,\,\, IQR\,\textsuperscript{\textbf{ii)}} & X: 0.23, Y: 0.14, Z: 0.11  & Y: 0.06, Z: 0.08 & Z\textsubscript{CL}: 0.1, Z\textsubscript{OL}: 0.1 & Y: 0.31, Z: 0.28 \\[0.15\normalbaselineskip]
   \hline
      & & & & \\[-2ex]
   Group delay\,\textsuperscript{\textbf{iii)}} ($f$~=~28\,Hz) / ms & Y: 9.9, Z: 9.5 & Y: 1.9, Z: 1.8 & Z\textsubscript{CL}: 1.2, Z\textsubscript{OL}: 1.5 & Y: 4.1, Z: 4.1 \\[0.15\normalbaselineskip]
   \hline
   & & & &  \\[-2ex]
   Settling time\,\textsuperscript{\textbf{iv)}} / ms &  &  &  &   \\[0.15\normalbaselineskip]
   \,\,\, Median ($\pm$~5\,\% criteria)  & 23.3 & 6.0\,\textsuperscript{\textbf{i)}} & Z\textsubscript{CL}: 2.8 & 7.3 \\[0.15\normalbaselineskip]
   \,\,\,\,\,\,\,\,\, IQR & 18.3\,\textsuperscript{\textbf{v)}} & 1.5 & Z\textsubscript{CL}: 0.1 & 0.5 \\[0.15\normalbaselineskip]
   \,\,\, Median ($\pm$~1\% criteria) & 75.3 & 51.3
   & Z\textsubscript{CL}: 3.3 & 22.9 \\[0.15\normalbaselineskip]
   \hline
    \end{tabular}
    \begin{tablenotes}
    \item[i)] \textsuperscript{\textbf{i)}} Dual axis, closed loop mode, software update (April 2025). X-axis is excluded here, as it is used solely for CAPE compensation. Its -3dB-bandwidth is below 10\,Hz and a median of about 4.7\,Hz.
    \item[ii)] \textsuperscript{\textbf{ii)}} IQR inter quartile range: includes 50\,\% of values, statistical outliers are excluded. 
    \item[iii)] \textsuperscript{\textbf{iii)}} Group delay value corresponds to a representative (typical) sensor within each system.
    \item[iv)] \textsuperscript{\textbf{iv)}} Limited sampling rates (1 and 1.5\,kHz) introduce an estimated uncertainty of $\approx \pm$1\,ms for single channel measurements.
    \item[v)] \textsuperscript{\textbf{v)}} Due to a bimodal distribution, the IQR not reliably represent the underlying variability.
   \end{tablenotes}
 \label{tab:results}
\end{table}

From Table\,\ref{tab:results}, it can be seen that the 3D Neuro-1 system (QuSpin Inc.) and the 2.5D HEDscan system (FieldLine Inc.) exhibit distinct inter-axis differences in median time delay values (X, Y, Z). These span the full range of observed delays at 28\,Hz, from less than 2\,ms to almost 10\,ms. The HEDscan (X-axis) is excluded from this comparison, as it is designed exclusively for CAPE compensation. Due to its low operational bandwidth ($f\tidx{-3dB} \approx 5$\,Hz, triaxial mode), it exhibits substantially longer delays, approximately 37\,ms at 1\,Hz. In triaxial mode, the bandwidth of the Y- and Z-channels is currently significantly reduced. Therefore, the Y- and Z-axis of the HEDscan system were operated in dual-axis mode during all measurements. The single-axis FLv2 system, operated in CL mode, exhibited the shortest median time delay ($M(t\tidx{d}) \approx 1.2$\,ms), reflecting its high $f\tidx{-3dB}$-bandwidth and optimized dynamic response. All three systems, Neuro-1, HEDscan, and FLv2, were operated in CL mode during testing. Compared to OL operation, CL operation typically provides improved dynamic range, broader bandwidth, enhanced operational stability, better linearity \cite{Alem.2023}, and facilitates CAPE suppression when applied on three orthogonal axis, although it may introduce a slight increase in noise. The intra-channel time delay variability, expressed as the interquartile range (IQR), was well controlled in all systems, as expected for high-quality multichannel platforms. However, manufacturer-specific design strategies led to notable differences: QuSpin targeted an IQR of less than 0.35\,ms for the dual-axis QZFM Gen.2 sensors and less than 0.55\,ms for the more intricate 3D Neuro-1 system. In contrast, the FieldLine systems (HEDscan and FLv2) achieved tighter intra-channel calibration, with IQR values below 0.1\,ms for the primary sensing axes. The FLv2 system was evaluated in both OL and CL configurations, highlighting the performance gains associated with closed-loop operation and frequency-dependent transfer function calibration. All systems demonstrated well-controlled time delay behavior in the MEG-relevant frequency range, exhibiting near-linear phase characteristics. Details are provided in the discussion chapter.

The next parameter listed in Table\,\ref{tab:results} is the typical group delay measured at $f=$\,28\,Hz. Variations in group delay as a function of frequency ($t_g\left(f\right)$) provide a direct indicator of signal integrity, and are discussed in more detail for the following results. In general, the results reflect the trends observed for time delay: both the Neuro-1 and QZFM Gen.2 from QuSpin exhibit significantly longer group delays at 28\,Hz compared to the FieldLine systems (HEDscan and FLv2). This indicates divergent optimization strategies between the manufacturers. Group delay spread above 20\,Hz is minimal and thus IQR is omitted from the table for all systems. The comparable behavior across the X-, Y-, and Z-axis enables the settling time of each MCS to be represented in Table\,\ref{tab:results} by a single median value and corresponding IQR. The reported values highlight a clear distinction between legacy and current-generation systems: the predecessor models, QZFM Gen.2 and FLv2, demonstrate significantly shorter settling times compared to their more complex successors. Among all systems, the single-axis FLv2 shows the fastest settling time, along with the narrowest intra-channel variability. Furthermore, the observed ratio between the $\pm5$\,\% and $\pm1$\,\% error band criteria suggests that the settling response cannot be accurately modeled by a single exponential decay, indicating a more complex dynamic behavior.

\begin{figure*}[t!]
    \centering           
    \includegraphics[width=1\textwidth]
    {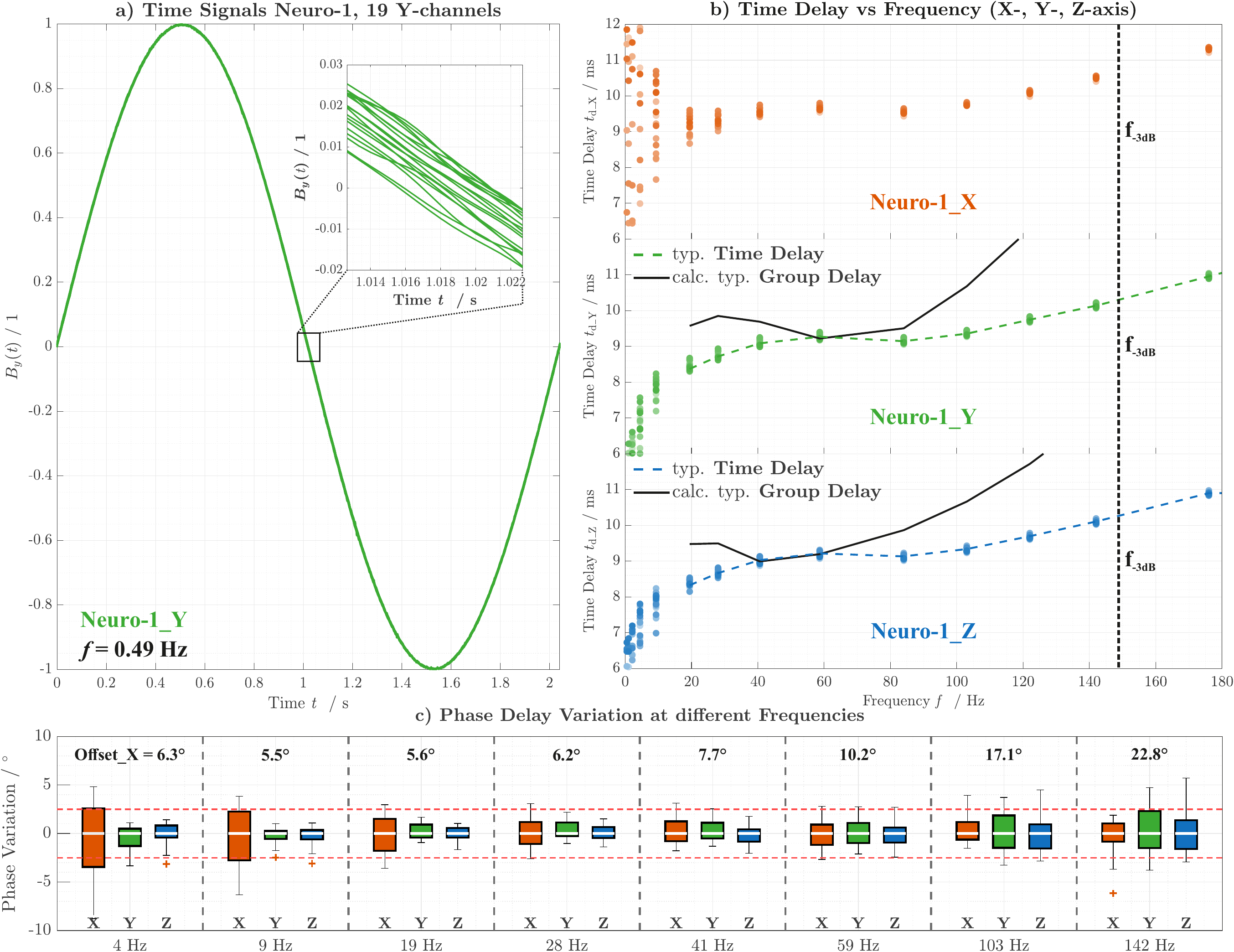}
    \caption{\justifying Neuro-1 results: a) Normalized time signal, overlap of 19 Y-channels with an applied test field of $f$ = 0.49\,Hz. The inset highlights the zero-crossing region for visualizing relative time delay variation between the sensors. b) Time delays in dependency of frequency over sensitive axes (X, Y, and Z). Data points represent measured delays to the applied test field of $B\tidx{exc}$ = 400\,pT; dashed lines indicate typical time delay characteristic; black lines show calculated group delay. Notably, the X-axis exhibits significant deviations in the low-frequency range; therefore, no typical time delay or group delay is calculated. c) Tukey-style boxplots illustrate the phase delay variation, defined as the interquartile range (IQR) relative to the median, across MEG-relevant frequency bands and axes (X, Y, Z). The median of X-channels shows a systematic phase offset relative to Y-, Z-channels, additionally annotated above each frequency group. Phase variation is mostly within red dashed lines at $\pm 2.5^{\circ}$.}
    \label{fig:synchronicity_qzfm}
\end{figure*}

Fig.\,\ref{fig:synchronicity_qzfm} summarizes key timing parameters of the Neuro-1 system as a function of frequency and additionally illustrates the effect of time delay in the time-domain: Fig.\,\ref{fig:synchronicity_qzfm}~a) shows one period of a 0.49\,Hz sinusoidal magnetic test signal, recorded simultaneously across 19 Y-channels of Neuro-1, is shown for illustrating the effect of time delay in the time domain. The inset highlights intra-channel time delay (green traces), visualized near the zero-crossing. Fig.\,\ref{fig:synchronicity_qzfm}~b) displays time delay versus frequency for the X-(copper-red), Y-(green), and Z-(blue) axes, recorded in response to a 400\,pT excitation. Dashed colored lines represent fitted time delay characteristics of the typical sensor, while solid lines indicate the corresponding typical group delay derived from typical time delay data. A steep digital filter limits the effective system bandwidth to approximately 150\,Hz, as shown by the black dashed line. The frequency dependence of the time delay exhibits a reduced spread within the frequency band of  20~–~140\,Hz, reflecting intentional manufacturer optimization to ensure uniform timing characteristics across the sensor array. Significant scatter is observed for the X-axis below 28\,Hz; hence, no typical time or group delay is provided for this axis. In contrast, the Y- and Z-axis data exhibit consistent statistical behavior across sensors. Fig.\,\ref{fig:synchronicity_qzfm}~c) presents Tukey-style box plots showing the phase delay variation (IQR relative to the median) across selected discrete frequencies. The Y- and Z-channels exhibit nearly identical medians, while X-channels show a systematic phase offset, annotated per frequency. Overall, phase variation remains below $\pm 2.5^\circ$, with smaller deviations at lower frequencies, indicating a well-calibrated system within the typical MEG frequency band.

\begin{figure*}[ht!]
    \centering           
    \includegraphics[width=1\textwidth]{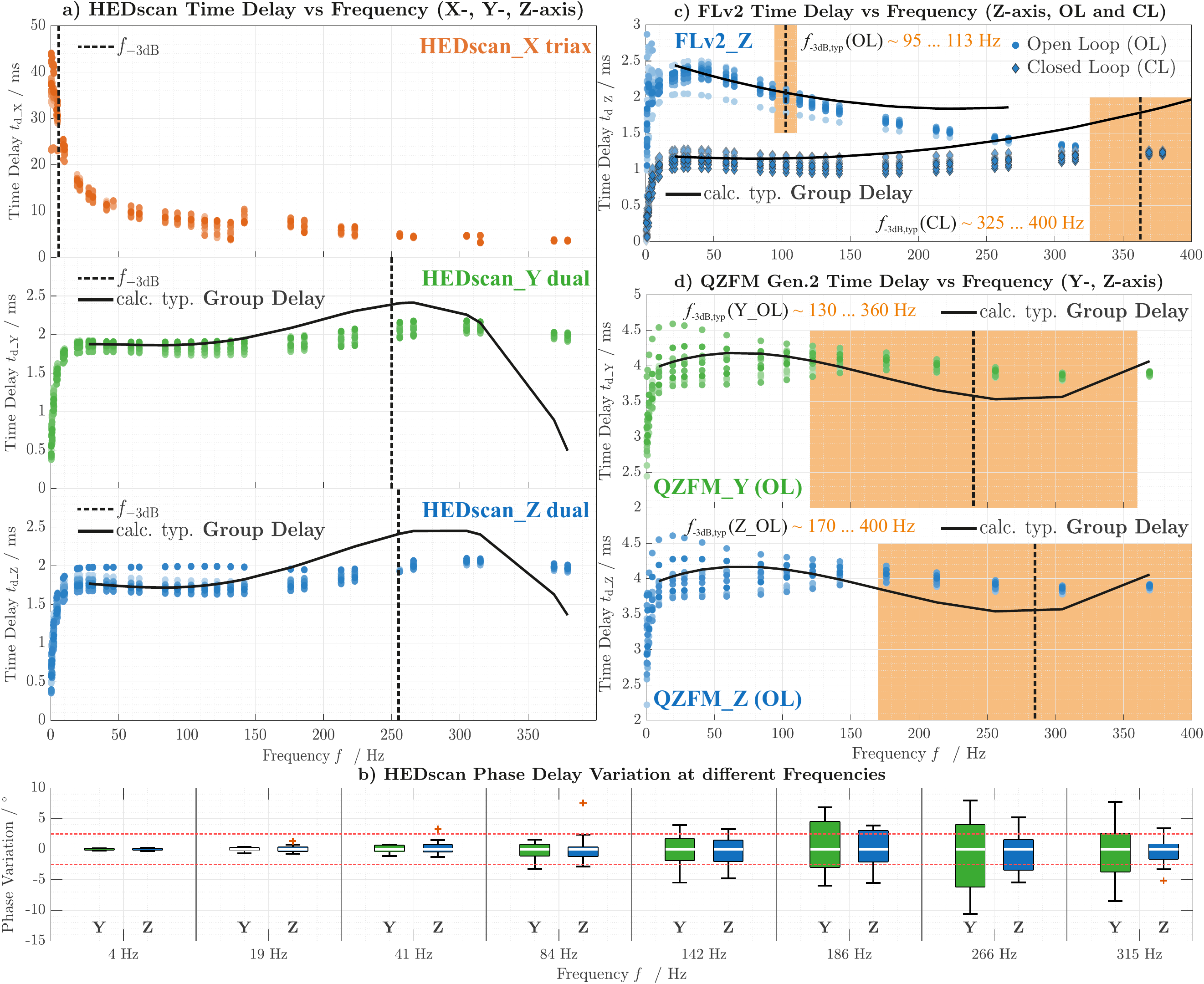}
    \caption{\justifying HEDscan, FLv2 and QZFM Gen.2 results, all for an applied test field of $B\tidx{exc}$ = 400\,pT. Data points represent measured delays to the applied test field; black lines show calculated group delay: a) Time delays of HEDscan in dependency of frequency and X-, Y-, and Z-axis. Notably, Y- and Z-axes are characterized in dual mode; X-axis has a narrow bandwidth, corresponding to much larger time delays and significant deviations in the low-frequency range; therefore, no typical group delay is calculated. b) Tukey-style boxplots illustrate the phase delay variation of HEDscan in dual mode (Y, Z), defined as the inter-quartile range (IQR) relative to the median, across different frequencies. Phase variation is mostly within red dashed lines at $\pm 2.5^{\circ}$. c) Time delays of FLv2, single-axis sensors, for two operational modes (Open Loop and Closed Loop), including measured -3dB-bandwidth range for both modes. d) Time delays of QZFM Gen.2, dual axis, open loop mode, with a large spread of supported -3dB bandwidth.}
    \label{fig:HEDscan_FLv2_QZFM}
\end{figure*}

Fig.\,\ref{fig:HEDscan_FLv2_QZFM} presents time and group delay comparisons for the remaining three systems: HEDscan (2.5D) and FLv2 (1D) from FieldLine Inc., and the Neuro-1 predecessor based on 16 individual QZFM Gen.2 (2D) sensors from QuSpin Inc. The HEDscan system is referred to as a 2.5D MCS because its X-axis, limited to a bandwidth of M($f\tidx{-3dB}$) = 4.7\,Hz, does not contribute to signal information but is used for CAPE compensation, only. This has been validated for quasi-static X-axis field variations up to $B\tidx{exc}=\pm$ 50\,nT. The comparison reveals distinct system design strategies: QuSpin pays attention to phase delay alignment within the frequency range of 20~-~140\,Hz, with reduced effort. In contrast, FieldLine focuses on achieving high signal integrity before A/D conversion by calibrating the entire sensor transfer function, encompassing amplitude, bandwidth, and phase, across the full operational bandwidth. However, an anti-aliasing filter necessary for the suppression of shadow signals is not incorporated (cf. Supplement).

In Fig.\,\ref{fig:HEDscan_FLv2_QZFM}~a), time delay versus frequency is shown for the HEDscan X-(copper-red), Y-(green), and Z-(blue) axis in CL mode, covering $f$ = 0.5~-~400\,Hz. FieldLine Inc. remotely calibrated the whole system in triaxial mode (X-axis) and dual-axis mode (Y-axis, Z-axis) during system maintenance for BMSR-2.1. Within this investigation, the performance of the Y- and Z-channels is shown only in dual mode calibration. The time delay characteristics are modeled based on the measurement results using a 4\textsuperscript{th}-order polynomial, $t\tidx{d}(f) = \sum_{n=0}^4 a_n f^n$, with only minor variability in the constant term $a_0$, likely due to calibration uncertainty. This time delay approximation simplifies the calculation of group delay (cf. Subsec.~\textit{Group Delay}). The corresponding typical group delay profiles for Y- and Z- are shown as solid black lines, with the typical system’s –3\,dB-cutoff-frequency ($f\tidx{-3dB}$) indicated by vertical dashed lines. The Y- and Z-axis behave consistently and are well-described by a typical delay and spread, indicating effective calibration. Group delay differences within a sensor remain significantly below the 1\,kHz sampling resolution, making further correction unnecessary. Notably, FieldLine’s calibration extends beyond timing parameters: the amplitude transfer function (not shown) is also individually adjusted, achieving a flatness better than $\pm$5\,\% across the 20~–~140\,Hz frequency range. 

Fig.\,\ref{fig:HEDscan_FLv2_QZFM}~b) shows Tukey-style box plots of the phase delay variation (IQR relative to the median) for the HEDscan system across selected frequencies. The Y- and Z-axis exhibit closely matched medians with IQRs consistently below $\pm 2.5^\circ$ up to 150\,Hz. Phase variation is particularly low at lower frequencies, indicating a well-calibrated, near-linear phase system. The corresponding time delay variation remains below 0.1\,ms for both axes. One Z-axis outlier shows slightly increased deviation, marked by individual points. 

\begin{figure*}[b!]
    \centering   
    \includegraphics[width=0.95\textwidth]
    {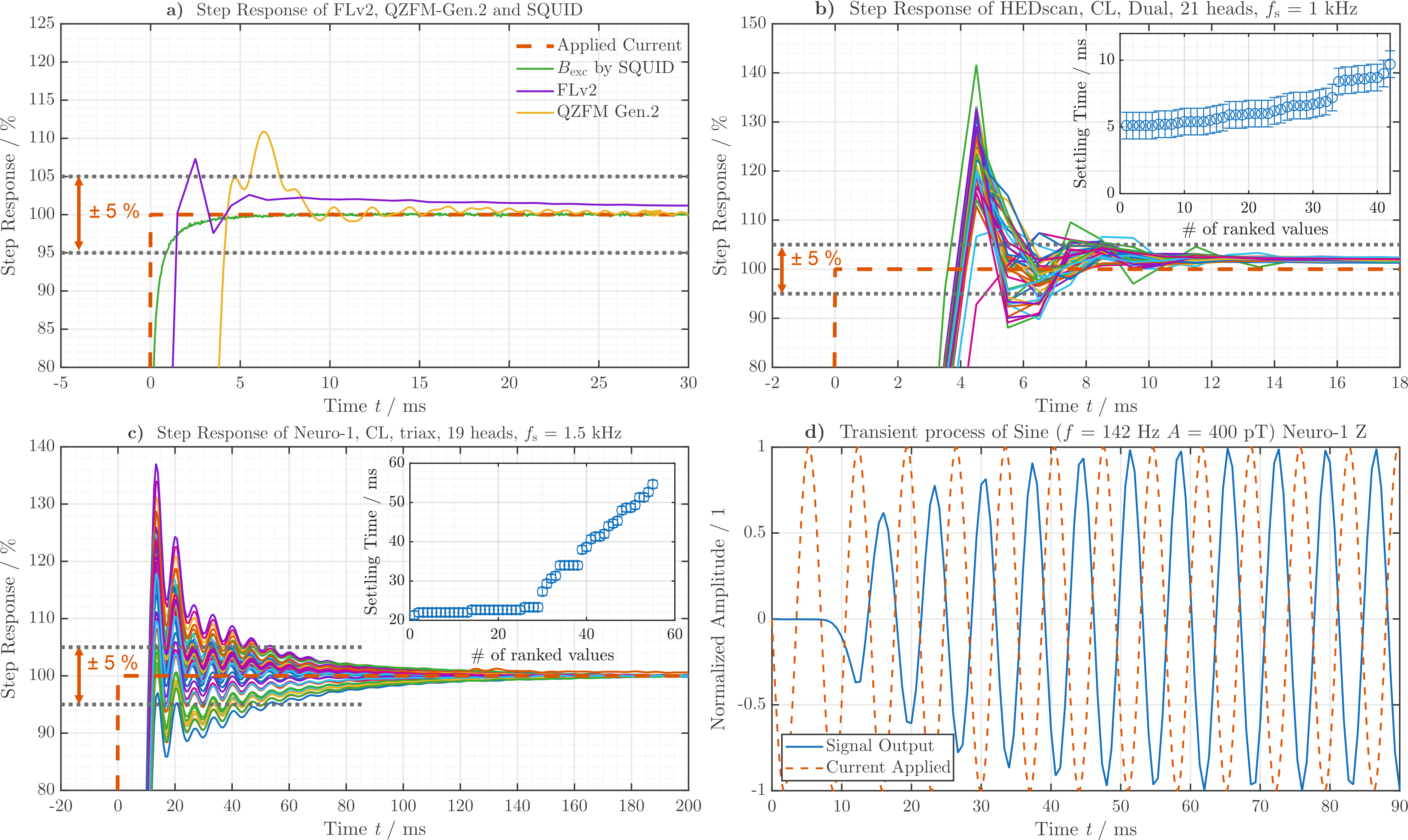}
    \caption{\justifying Settling time responses of the investigated multichannel SERF-OPM systems: The applied current step is shown as a dashed red line, initiated at $t$ =\,0\,ms. Signal amplitudes are normalized to the steady-state value and expressed as a percentage. Settling time is defined as the interval between $t = 0$\,ms and the point where the response remains within $\pm$~5\,\% of the target amplitude, marked by black dotted lines. a) SQUID measurement (green) characterizing the step function of the applied field. Example responses from QZFM Gen.2 (OL, yellow, $\Delta B$(QZFM) = 400\,pT) and FLv2 (CL, lilac, $\Delta B$(FLv2) = 1000\,pT) are overlaid. b) Step responses of 21 HEDscan heads in closed-loop, dual-axis mode $\Delta B$(HEDscan) = 1000\,pT. The inset shows the distribution of settling times using the $\pm$5\,\% criterion, with an estimated uncertainty of $\pm~1$\,ms. c) Step responses of 19 Neuro-1 heads in closed-loop, triaxial mode, yielding 57 traces (3 x 19 curves) for a 1000\,pT step height. A bimodal distribution is observed. d) Neuro-1 Z-channel response to a 142\,Hz, 400\,pT sine wave, illustrating the settling time introduced by the system’s dominant low-pass behavior. Approximately 50\,ms are required for the signal to reach its steady-state amplitude, highlighting the required settling before accurate amplitude measurement is possible.} 
    \label{fig:settling}
\end{figure*}

Fig.\,\ref{fig:HEDscan_FLv2_QZFM}~c) presents the time delay characteristics of the FLv2 single-axis MCS in CL (black diamonds) and OL (blue dots) modes. As expected, time delays in CL mode are significantly lower within the calibrated range of 10 to 300\,Hz, reflecting improvements in bandwidth and dynamic range (not shown). Both time delay curves exhibit a second-order frequency dependence, resulting in a more substantial deviation at frequencies < 20\,Hz, as indicated by the group delay (black line). The CL system exhibits smooth group delay behavior from 20\,Hz to 300,Hz, closely following the time delay with only minor deviations above 200\,Hz. Although the typical cut-off point for $f\tidx{-3dB}$ is around 360\,Hz, the 1\,kHz sampling rate ($f\tidx{s}$) limits the resolution.

Fig.\,\ref{fig:HEDscan_FLv2_QZFM}~d) shows the time delay characteristics of 13 individual 2D QZFM Gen.2 sensors (QuSpin Inc.) operating in OL mode, mounted on the DALAC and used in dual-axis mode. Y- and Z-channels are shown as green and blue dots, respectively. These sensors exhibit relatively short typical delay times $t\tidx{d,typ}$ = 4\,ms, large $f\tidx{-3dB}$ cutoff frequency, and well-controlled group delay behavior, especially when compared to the newer Neuro-1 system. However, the inter-sensor variation is considerable, indicating that these sensors were not originally designed to meet the performance uniformity required for proper multichannel system operation.\\\vspace{-3mm}

Fig.\,\ref{fig:settling} summarizes the settling time characteristics of the investigated multichannel SERF-OPM systems, as determined by their response to a magnetic field step change (DC step). Settling time reflects the combined influence of time delay, slew rate, filter characteristics, and the natural transfer characteristics of the DuT. If the slew rate were the limiting factor, settling time would increase with the step amplitude; however, no such dependency was observed in the measurements. For QZFM Gen.2 sensors operating in OL mode, step amplitudes between 100\,pT to 400\,pT were applied to remain within the linear range. All other systems were tested in CL mode using step amplitudes between 100\,pT to 1000\,pT. Measurement accuracy is influenced by both the rise time of the applied magnetic step and the sampling rate of the data acquisition. The characteristic of the applied step $\Delta B$ is measured by an SQUID with a sampling rate of 32\,kHz and presented in Fig.\,\ref{fig:settling}~a) to confirm that the test field $B\tidx{exc}$ is fast enough to evaluate the DuTs (cf. green curve). Also shown are representative step responses from QZFM Gen.2 (yellow, $f\tidx{s}$ = 8\,kHz) and FLv2 (lilac, $f\tidx{s}$ = 1\,kHz). The low sampling rates of 1.5\,kHz (Neuro-1) and 1\,kHz (HEDscan, FLv2) limit the temporal resolution, reducing the precision and reproducibility of settling time estimates at the single-channel level. To account for subsampling effects, an uncertainty of approximately $\pm$1\,ms is assumed. Therefore, robust statistical measures, median and inter-quartile range (IQR), are used to characterize the settling time of the MCSs. Fig.\,\ref{fig:settling}~b) shows a representative measurement of $t\tidx{set}$ for the HEDscan ensemble. The distribution in the inset supports the use of median and IQR as appropriate metrics for quantifying settling time, despite limited single-channel reproducibility caused by sub-sampling. For the Neuro-1 system, all channels exhibit natural resonance with approximately 150~–~155\,Hz, as shown in Fig.\,\ref{fig:settling}~c). Approximately half of the channels show relatively short and consistent settling times around 23\,ms, while the remainder show a wider spread, resulting in a bimodal distribution (see inset). The observed response is a consequence of the steep low-pass filtering implemented in the system. Fig.\,\ref{fig:settling}~d) illustrates the filter response of a Neuro-1 Z-channel to a 142\,Hz, 400\,pT sine wave (amplitude normalized shown). The Neuro-1 channel requires approximately 50\,ms to reach its steady-state amplitude, underscoring the delay introduced by the system’s low-pass filtering.

\section*{Discussion} 
\label{sec:discussion}
QuSpin Inc.\ and FieldLine Inc.\ have focused their system development on stable, low-intrinsic-noise, two- to three-dimensional SERF-OPMs tailored for nearly interruption-free MEG recordings inside MSRs with probate shielding. Both novel systems implement three-dimensional FLL compensation at the vapor cell, actively canceling field changes caused by sensor motion within a static residual field. Optimal performance of an MCS is achieved only when all channels exhibit nearly identical characteristics, namely: gain, amplitude response, phase (time-delay) response versus frequency, directivity, and linear range. QuSpin and FieldLine adopted different strategies to minimize channel-to-channel parameter variation.

\begin{itemize}
   \item \textbf{Neuro-1:} A digital uniform time-delay equalization approach introduces an additional individual time delay per channel, minimizing delay spreading above 20\,Hz at the cost of a typical $\approx$ 9.3\,ms total time-delay between 20\,Hz and 100\,Hz. A steep low-pass filter at $\approx$ 150 Hz is acting as an anti-aliasing filter. The delay requires user-based correction when synchronizing the data with external modalities, such as EEG or stimulation trigger events. Other necessary error correction and distortion cancellation methods are planned to be performed after the DAQ system in the digital domain by the end-users, applying the extensive available stochastic digital signal processing toolboxes. QuSpin and its collaborators have also developed two complementary calibration schemes to improve channel matching. A 3D magnetic field generated by 3D multi coil panels -- or, alternatively, by a "HALO"-hardware extension board \cite{QuSpin.2024} attached to the helmet and comprising 16 spatially displaced dipole sources -- is applied during system initialization to determine each channel’s gain, position, and orientation at a single reference frequency \cite{Hill.2025}. Below 20\,Hz the time delay spread increases significantly. At 0.5\,Hz the adjustment range of approx. 6\,ms is detectable. The X channels show a systematic, frequency-dependent phase offset up to 23$^{\circ}$ relative to the Y- and Z-channels and significant stronger time delay variations below 20\,Hz. Thus, X channels are inefficient for electronic gradiometer-based distortion cancellation. Notably, the group delay peaks in the 20-30\,Hz range, well within the $\beta$-band of MEG, accompanied by substantial spread below 30\,Hz caused by inter-channel time-delay differences. This spread introduces additional non-linearities in distortion-cancellation algorithms and limits the achievable SNR. Above 60\,Hz the group delay rises continuously, reaching up to 14\,ms near the 150\,Hz cutoff frequency, while exhibiting negligible intra-channel spread above 80\,Hz. The group- and time-delay variations are adequately compensated and exert insignificant influence on today’s low-bandwidth, statistically minimal-variance MEG analysis approaches. Long settling times, however, restrict access to the early, information-rich phases of single-event responses, especially in high-rate stimulation protocols or when capturing rapid physiological transients. Neuro-1 channels require at least 22\,ms to stabilize after a transient field change. The 3~x~19 channels display a distinct bimodal settling-time distribution, necessitating about 100\,ms of data clearance after a large disturbance to ensure the slowest channel has surely settled, cf. Fig.\,\ref{fig:settling}~c) and inset. When using TCP/IP for data access, the data is sent to the PC every 80\,ms. An external FPGA module is currently under development to support real-time AI and BCI applications \cite{quspin2024neuro1}; this was further confirmed through direct vendor communication on August 21, 2025. Worth mentioning, the highly miniaturized ECU for the three-axis SERF-OPM incorporates the ADC and serial digital interface and is electrically decoupled from the DAQ unit, thereby reducing cable feed-through in MSRs. It is sufficiently compact to be worn as a backpack inside the MSR for moving proband experiments.

   \item \textbf{HEDscan} (iterim version hedscan-0.10.12-g64fb6e7, dual mode): A remote, on-site matching of individual transfer-function yields a short time delay of $\leq$ 2\,ms, with an IQR max. $\leq$ 0.1\,ms across 0.5–200\,Hz, enabling inherent gradiometer-based distortion cancellation that increases SNR and improves dipole-localization error (DLE) in MEG applications, cf. Fig.\,\ref{fig:HEDscan_FLv2_QZFM}~a. In the frequency range from 20 to 250\,Hz the group delay change versus frequency is below 7\,$\upmu$s/Hz. The low sampling rate of $f\tidx{s}=$ 1\,kSa/s compromises the timing performance for stimulation-evoked analysis. Together with a missing anti-aliasing filter (cf. \textit{Supplement}) to suppress mixing effects caused by YZ-channel demodulation frequency $f\tidx{m}$ = 500\,Hz, the time-domain signal integrity could be affected significantly. System users can overcome this issue by filtering out main aliasing components with small bandwidth analysis or steep 250\,Hz filters, resulting in increased time delay and longer settling times.  In the raw data, a reasonable median settling time of 6\,ms (on $\pm$5\,\%  criteria) is achieved in dual-axis mode, cf. Fig.\,\ref{fig:settling}~b). Data acquisition, transfer, and control are engineered for an end-to-end system to fulfill low overall system reaction times (<~50\,ms, including robotic feedback) in BCI applications. The manufacturer provides a Real-time Data Access (RDA) to accomplish this task. The individually channel matching approach simplifies post-processing, supports precise physical parameter estimation, and enhances gradiometric noise rejection. Notably wide bandwidth in CL operation, achieved via high feedback gain, substantially extends both dynamic and linear ranges, as demonstrated for FLv2 in Fig.\,\ref{fig:HEDscan_FLv2_QZFM}~c). Consequently, the system can operate in moderately shielded rooms. The triaxial mode has not been thoroughly investigated yet. It offers additional CAPE error compensation but introduces further distortions and reduces bandwidth: X-channels exhibit a median bandwidth of only $f\tidx{-3dB} \approx$ 5\,Hz, while Y- and Z-channel bandwidth are currently limited by a 250\,Hz notch filter used for a 2$^{nd}$-carrier-suppression in triaxial operation. 

   \item \textbf{FLv2 and QZFM Gen.2}: The predecessor MCS from QuSpin and FieldLine still have their specific application areas at the cost of CAPE compensation. FLv2 stands out with the larger bandwidth $f\tidx{-3dB} \approx$ 360\,Hz in CL mode, the smallest time delay $t\tidx{d}$ of 1.2\,ms with a low spread IQR < 100\,$\upmu$s below 400\,Hz and a steep low-pass filter at about $f\tidx{s}/2 $ = 500\,Hz. Furthermore, the gain matching versus frequency ($\Delta G\tidx{flat}<\pm$ 2.5\,\% up to 140\,Hz, not shown here) is the best within the investigated MCS. The group delay $t\tidx{g}$~=~1.2\,ms is flat up to 200\,Hz. This system demonstrates the shortest settling time $t\tidx{set} \approx$~2.8\,ms but at a low sampling rate of only 1\,kSa/s, which increases the uncertainty of the result. All this makes the FLv2 a good choice for fundamental science investigation demanding precise magnetic field measurements with a high degree of signal integrity. The QZFM Gen.2 OPMs remain popular for broadband, low-intrinsic-noise investigations, extending well above the $f\tidx{-3dB}$ bandwidth. By picking large bandwidth sensors, special investigations can be performed up to 500\,Hz with user selectable sampling rates and probate increase in intrinsic noise \cite{Kruse.2025, Nordenstrom.2025}. Nevertheless, the limited bandwidth reduces the diagnostic information obtainable from broadband neurological recordings and constrains the precise determination of physiological conduction latencies when compared with the bandwidth available in the clinical electrical gold standards of Electromyography (EMG) and Electroneurography (ENG) \cite{Elzenheimer.2021, Kimura.2013}. Achieving high precision in fundamental research requires detailed characterization of these OPMs, a straightforward task for a few sensors but laborious for MCS. The OL mode QZFM Gen.2 OPMs measured in this investigation present an IQR of 0.31\,ms (Y-channel) at 28\,Hz, which is a substantially wider spread than the corresponding Neuro-1 sensors, but with a shorter median time delay of 4\,ms for Y- and Z-channels. The significant spread of the $f\tidx{-3dB}$ bandwidth, depicted in Fig.\,\ref{fig:HEDscan_FLv2_QZFM}~d) with brown background, visualizes the substantial variability of the OPM head and the complex sensor parameters interaction in OL mode. Furthermore, a basic characterization of the QZFM Gen.3 (OL, triaxial) sensors can be found in \cite{Nordenstrom.2024}.
\end{itemize}

In Magnetoencephalography (MEG), sub-millisecond inter-channel time synchronization is essential for high-resolution source localization, minimal localization bias, and robust interference suppression \cite{Munoz.2009, Long.2024}. The four systems tested meet this requirement in the 20-140\,Hz frequency band. They also exhibit an acceptable group delay, $t\tidx{g}$ < 15\,ms, which preserves waveform integrity for common brain investigations. However, the relatively low sampling rates of these state-of-the-art MCS introduce additional timing uncertainty in triggered, evoked, or other fast-transient applications. Battery-powered, wireless systems, as demonstrated by Cheng et.al. \cite{Cheng.2024}, could increase flexibility for movement experiments and reduce power-line distortions. In the context of widely utilized MEG-/EEG-systems, correlation-based methodologies or minimum variance approaches (like linearly constrained minimum variance (LCMV) beamforming) are commonly employed to mitigate the impact of environmental interferences and distortions. They are frequently implemented in the context of Statistical Parametric Mapping (SPM) to minimize signal power through the utilization of linear regression techniques and minimum-norm methods \cite{Friston.2007}. Although they are effective over a broad range of parameters and do not require explicit group-delay compensation, this can cause a substantial loss of medical diagnostic information.

Differences in delay among sensing axes are particularly detrimental for source analysis methods based on LCMV beamforming. Beamformers attenuate signal components that are not aligned with the selected signal-space vector, typically defined by the forward model of a dipole at a given location and orientation. Consequently, even minor deviations in gain or time delay can lead to pronounced attenuation of beamformer source estimates. Simulation studies are required to quantify the relation between timing performance and DLE. This challenge is especially pronounced in OPM-MEG, where the magnetometer response exhibits higher susceptibility to artifacts from environmental and biological sources (e.g., MCG, eye movements) compared to SQUID gradiometers. Furthermore, the construction of gradiometric responses through subtraction of primary and reference sensors requires matching of both gain and phase characteristics \cite{Vrba.2001biomag, Eswaran.2000, Vrba.2001, Vrba.2002}.

Accurate knowledge and effective minimization of time delays are especially critical for BCI applications \cite{Soekadar.2023} and closed-loop neuromodulation \cite{Nasr.2022}, where precise interpretation of real-time signals is critical. Even small variations in delay can degrade phase alignment, reduce decoding accuracy, and impair the precision of brain signal classification, particularly in high-frequency or fast-changing neural dynamics. For such applications, sub-millisecond synchronicity and stable group delay become not just desirable but essential. Therefore, timing performance must be characterized and optimized with the same rigor as signal quality, bandwidth, and noise suppression when evaluating multichannel OPM systems. 

Future investigations should systematically address: a) the angular dependence of sensor gain and time delay (directivity); b) real-time performance to confirm suitability for the projected expansion of BCI applications; c) inter-head crosstalk in multichannel systems, which may impose significant constraints; and d) sensor stability and reliability, a critical determinant for clinical and long-term scientific investigations.
 
\section*{Conclusion}
This work presents the first cross-platform MCS assessment of timing performance metrics in commercial SERF-OPM multichannel systems, providing novel insights into their characteristics and limitations. The state-of-the-art MCSs Neuro-1 (QuSpin Inc., Louisville, CO, USA) and HEDscan (FieldLine Inc., Boulder, CO, USA) provide CAPE compensation by integrating a three-axis Flux-Locked-Loop mode, and distinct strategies were developed to effectively minimize channel-to-channel parameter variation. Using a digital time-delay equalization (Neuro-1, QuSpin) versus per-channel transfer-function matching (HEDscan, FieldLine) leads to pronounced differences in time delays, group delays, and settling-time dynamics. Transfer-function matching reduces the effort required for post-processing, including processing time. The four systems under test meet the basic phase-alignment requirements for common MEG within the 20 to 140\,Hz frequency range. Additionally, they exhibit an appropriate group delay $t\tidx{g}$ < 15\,ms to ensure waveform integrity. With an intra-channel time delay spreads below 1\,ms, they are entering the region where time delay is not limiting the Common-Mode-Rejection-Ratio (CMRR) of software gradiometers and global feedback systems. The implementation of CAPE compensation and delay equalization reduces usable bandwidth to approximately 150\,Hz for Neuro-1 and increases the settling time. The HEDscan maintains a wider pass-band range ($\approx$ 250\,Hz), and faster settling time $t\tidx{set}$~<\,10\,ms in the dual-axes mode, at the cost of CAPE effects, restricted subject/device movement, and alias effects due to a missing steep anti-aliasing filter. Until manufacturers provide comprehensive and standardized specification sheets, subsequent studies must characterize the non-ideal behavior of SERF-OPM multichannel systems. Furthermore, simulation studies are required to evaluate the impact of imperfect sensor responses on typical biomagnetic signals before OPM systems can be reliably considered for clinical applications.

\section*{Acknowledgment}
    \addcontentsline{toc}{section}{Acknowledgment}
     This research was funded by the German Research Foundation (DFG) through Projects: \textit{Z2} of the Collaborative Research Center 1261 “\textit{Magnetoelectric Sensors: From Composite Materials to Biomagnetic Diagnostics}” (Project-ID: 286471992) and "\textit{Identifying Circuits Dysfunctions in Schizophrenia with Optically-Pumped Magnetometers (OPM): A combined OPM-MEG/EEG and Computational Modelling Study}" (Project-ID 460785001), the Federal Ministry of Research, Technology and Space (BMFTR) with QHMI and QHMI2 (Project-ID: 03ZU1110DD, 03ZU2110FC), NeuroQ (Project-ID: 13N16486), and the Einstein Foundation Berlin. We express our sincere appreciation to several individuals for their invaluable contributions to our work. We thank F. Ptach for constructing, F. L\"oser and the specialized workshop at PTB (Berlin) for computer-aided design and non-magnetic manufacturing of the DALAC. Furthermore, we thank the OPM team at PTB, for the supporting discussions. We are also grateful to T. Sander for organizing the QuSpin systems and for his insightful SERF-OPM discussions. Open Access funding is enabled and organized by Project DEAL. 

\bibliography{references}
\newpage

\section*{Supplement}
\begin{figure*}[h!]
    \centering   
    \includegraphics[width=1\textwidth]
    {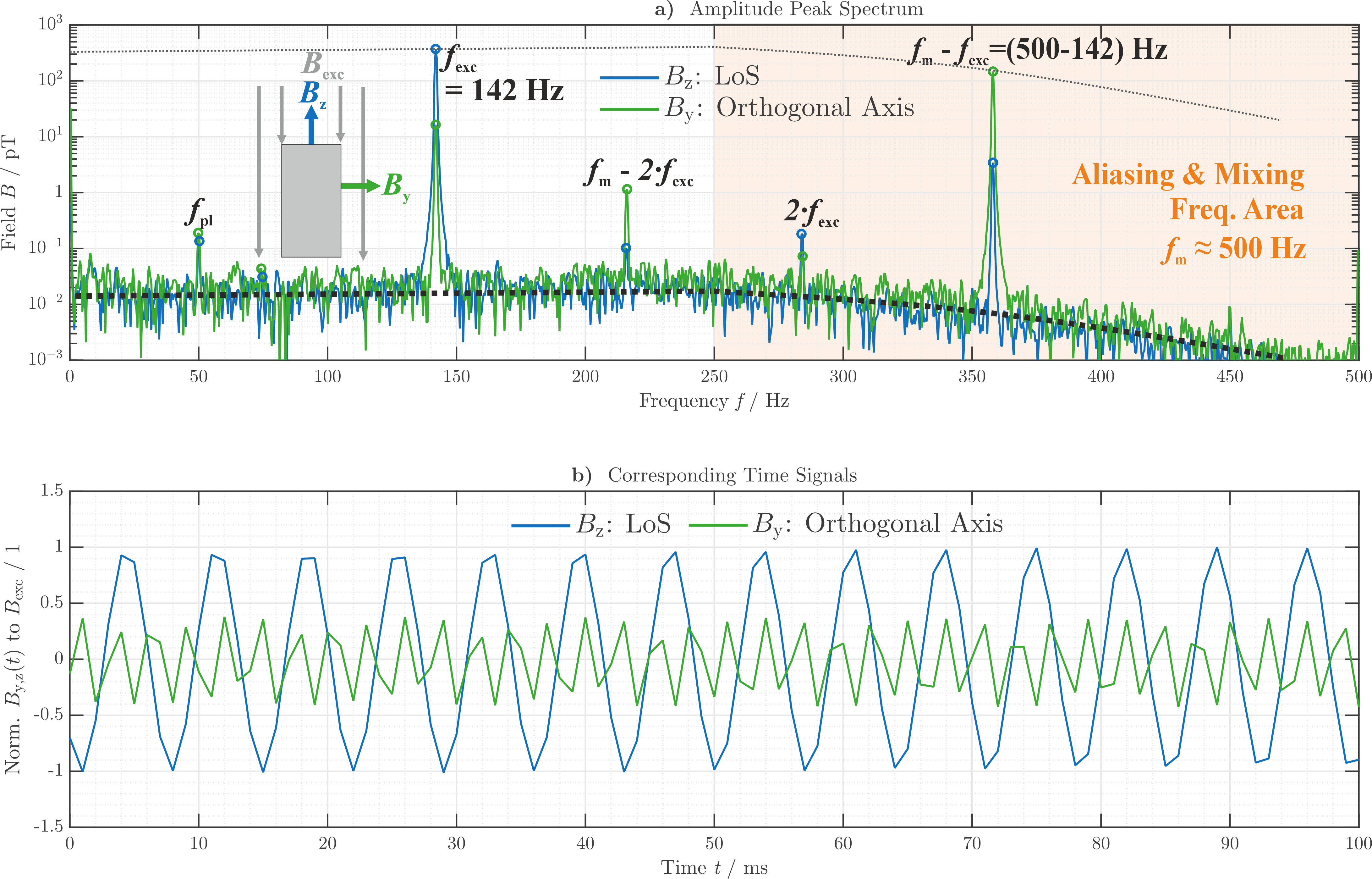}
    \caption{\justifying a) Amplitude spectrum calculated from a recording of a HEDscan sensor with an applied sinusoidal excitation of $B\tidx{exc} = 400$\,pT at $f = 142$\,Hz, measured in dual-axis mode along the testfield (Z-channel, blue) and the orthogonal, non-LoS to testfield (Y-channel, green). Ring markers indicate amplitudes at specific frequencies $f_m \pm n f\tidx{exc}$ or at $f\tidx{pl}$ in both channels. The appearance of second-order harmonics indicates non-linear sensor behavior. However, the resulting modulation products are not suppressed due to the absence of an anti-aliasing filter. b) Corresponding time-domain signals of the Z- and Y-channels, normalized to $B\tidx{exc}$, with $f\tidx{s}$ = 1\,kHz The Z-channel shows a sinusoidal response at the excitation frequency, while the Y-channel exhibits orthogonal non-LoS coupling and distortion effects.} 
    \label{fig:supplement}
\end{figure*}

During measurements of the HEDscan performance, unexpected frequency components were observed in the spectral domain, necessitating further consideration. Although not within the scope of this publication, these effects should be acknowledged. This effect became evident when a dedicated mono-frequency magnetic test signal was applied. Fig.\,\ref{fig:supplement}~a) shows amplitude spectra of a typical HEDscan sensor with a sinusoidal excitation field of $B\tidx{exc} = 400$\,pT at $f\tidx{exc} = 142$\,Hz, operated in 2D-CL dual-axis mode. The solid blue curve represents the measured field by the LoS Z-channel, while the solid green curve corresponds to the signal picked up by the orthogonal, non-LoS Y-channel. Ring markers indicate amplitudes at characteristic frequencies ($f_m \pm n f\tidx{exc}$) or at $f\tidx{pl}$, which represent power-line interference components. The underlying noise floor is consistent with the expected second-order CL low-pass characteristic.\\

The main LoS spectral peak appears as expected at 142\,Hz and is accompanied by a sensor-generated second harmonic at $2\cdot f\tidx{exc}$. In contrast, the other channel exhibits a shifted component at $f\tidx{m} - f\tidx{exc}$, where the modulation frequency is $f\tidx{m} = 500$\,Hz, as well as additional harmonic distortion products. The residual $f\tidx{exc}$ component observed in the orthogonal $B\tidx{y}$ channel originates from cross-coupling between the two channels of the sensor, while the complementary component at $f\tidx{m} - f\tidx{exc}$ in the $B\tidx{z}$ channel likewise arises from cross-coupling of the $B\tidx{y}$ field. This behavior reflects a characteristic of the special demodulation scheme utilized in this OPM scheme to separate Z- and Y-channel in a sensor that is intrinsically sensitive to both directions. Similar results are observed when the excitation field is randomly aligned, e.g., Y-channel in LoS, and with the most significant effect for 45$^\circ$ orientation of Z- and Y- channels (not shown). Fig.\,\ref{fig:supplement}~b) shows the normalized time-domain signals of the LoS channel (blue) and the orthogonal non-LoS channel (green). Ideally, the applied test signal should be detected exclusively by the LoS channel, which is clearly not the case. However, the presence of a measurable response in the non-LoS axis highlights residual cross-axis sensitivity of the sensor, which can compromise signal integrity and must be considered during analysis. The deviation from an ideal sinusoidal waveform is primarily caused by the sampling rate $f\tidx{s}$ and related interference, while harmonic distortions from the sensor itself contribute only to a lesser extent. These artifacts arise from the overall crosstalk between the Y- and Z-channels. Harmonic products of non-LoS signals may fold into the LoS passband even after post-processing filters, e.g., $f\tidx{m} - 2\cdot f\tidx{exc}$ when $f\tidx{exc}<250$\,Hz). For example, at $f\tidx{exc} = 84$\,Hz, a third-order distortion arising from the non-linear OPM transfer curve is clearly visible, producing components at $3\cdot f\tidx{exc} = 252$\,Hz and at $f\tidx{m} - 3\cdot f\tidx{exc} = 248$\,Hz.\\\vspace{-3mm}

The primary obstacle to practical use is the lack of detailed specifications from manufacturers that describe these sensor characteristics. Without this knowledge, users risk misinterpreting measurements due to artifacts originating from the OPM MCS rather than from a biomagnetic source itself. Awareness of these effects enables the development of countermeasures, while manufacturers can support users with adjustment procedures, software tools, and detailed guidelines. In addition, users must account for folding products arising from harmonics of non-LoS (orthogonal-axis) signals, which can fall into the bandwidth of interest and severely compromise source vector reconstruction (e.g., in LCMV beamforming with distortion cancellation) or produce dipole localization errors. More broadly, this modulation scheme complicates any gradiometer-based distortion cancellation method, since the time-domain signal is inevitably influenced by two field directions, with the impact depending on the sensor alignment. Furthermore, cross-coupling between channels of a single head is an inherent sensor characteristic and can vary significantly between heads. Inter-sensor crosstalk has not been considered here thus far.
\end{document}